%% file: 0_main.tex
\documentclass[journal,twoside,web]{ieeecolor}
\usepackage{jsen}
\usepackage{cite}
\usepackage{amsmath,amssymb,amsfonts}
\usepackage{algorithmic}
\usepackage{graphicx}
\usepackage{textcomp}
\usepackage{wrapfig}
\usepackage{multirow}
\usepackage{comment}
\usepackage{orcidlink}

\usepackage{color}
\usepackage{footnote} % footnotes!
\usepackage{hyperref}
\usepackage[]{siunitx}
\DeclareSIUnit{\degree}{°}
\DeclareSIUnit{\deg}{deg}
\DeclareSIUnit{\nothing}{\relax}
\DeclareSIUnit\pixel{px}
\DeclareSIUnit{\op}{Op}
\DeclareSIUnit{\mac}{MAC}
\DeclareSIUnit{\fps}{frame/s}
\DeclareSIUnit{\round}{round}

\usepackage{tabularray}
\usepackage{array}
\usepackage{booktabs} %\toprule, \midrule, \bottomrule
\usepackage{makecell}
\usepackage{threeparttable}
\usepackage{arydshln}

\usepackage{pifont} % for \cmark and \xmark if not already defined
\def\BibTeX{{\rm B\kern-.05em{\sc i\kern-.025em b}\kern-.08em
    T\kern-.1667em\lower.7ex\hbox{E}\kern-.125emX}}
\definecolor{abstractbg}{rgb}{0.89804,0.89804,0.89804}
\usepackage{eso-pic}
\usepackage{url}
\AddToShipoutPictureBG*{
  \AtPageUpperLeft{%
    \put(0,-30){\raisebox{-15pt}{\makebox[\paperwidth]{\begin{minipage}{21cm}\centering
      \textcolor{gray}{This paper has been accepted for publication in the IEEE Sensors Journal \copyright 2026 IEEE.\\DOI: \href{https://doi.org/10.1109/JSEN.2026.3713809}{10.1109/JSEN.2026.3713809}} 
    \end{minipage}}}}%
  }
  \AtPageLowerLeft{%
    \raisebox{20pt}{\makebox[\paperwidth]{\begin{minipage}{21cm}\centering
      \textcolor{gray}{\scriptsize \copyright 2026 IEEE.  Personal use of this material is permitted.  Permission from IEEE must be obtained for all other uses, in any current or future media, including reprinting/republishing this material for advertising or promotional purposes, creating new collective works, for resale or redistribution to servers or lists, or reuse of any copyrighted component of this work in other works.
      }
    \end{minipage}}}%
  }
}
\usepackage{fancyhdr}% Override the IEEE first-page header
\fancypagestyle{titlepagestyle}{
    \fancyhf{}
    \fancyhead[R]{\thepage}

}
\makeatletter
\long\def\IEEEtitleabstractindextext#1{%
  \def\@IEEEtitleabstractindextext{%
    \parbox{\textwidth}{%
      \bigskip\par
      #1
      \bigskip\par
    }%
  }%
}

\makeatletter
\long\def\IEEEtitleabstractindextext#1{%
  \def\@IEEEtitleabstractindextext{%
    \parbox{\textwidth}{#1}%
  }%
}
\makeatother

\title{\textcolor{black}{Thinking Fast, Thinking Slow: Adaptive Multimodal Transformer-based Sensor Fusion for Depth Estimation on Ultra-low-power MCUs}}
\author{
Luca Crupi \orcidlink{0009-0004-1391-8520}, 
Lorenzo Lamberti\orcidlink{0000-0003-1659-618X}, 
Giovanni Badaracco, 
Daniele Allegri\orcidlink{0000-0001-8762-792X}, 
Alessandro Giusti\orcidlink{0000-0003-1240-0768}, 
and Daniele Palossi\orcidlink{0000-0003-4487-0836}
\thanks{This work was partially supported by the SNSF RoboMix2 project (grant nb. 10004854).}
\thanks{Luca Crupi, Lorenzo Lamberti, Alessandro Giusti, and Daniele Palossi are with Istituto Dalle Molle di Studi sull’Intelligenza Artificiale (IDSIA), Scuola Universitaria Professionale della Svizzera Italiana (SUPSI), via La Santa 1, Lugano, 6900, Switzerland (e-mail: name.surname@supsi.ch).}
\thanks{Giovanni Badaracco and Daniele Allegri are with Istituto Sistemi e Elettronica Applicata (ISEA), Scuola Universitaria Professionale della Svizzera Italiana (SUPSI), via La Santa 1, Lugano, 6900, Switzerland (e-mail: name.surname@supsi.ch).}
\thanks{Lorenzo Lamberti and Daniele Palossi are also with the Integrated Systems Laboratory (IIS), Gloriastrasse 35, ETH Z\"urich, Z\"urich, 8092, Switzerland (e-mail: llamberti@iis.ee.ethz.ch).}}

\begin{document}

\IEEEtitleabstractindextext{%
\fcolorbox{abstractbg}{abstractbg}{%
\begin{minipage}{\textwidth}%
\begin{wrapfigure}[23]{r}{3in}%
\includegraphics[width=3in]{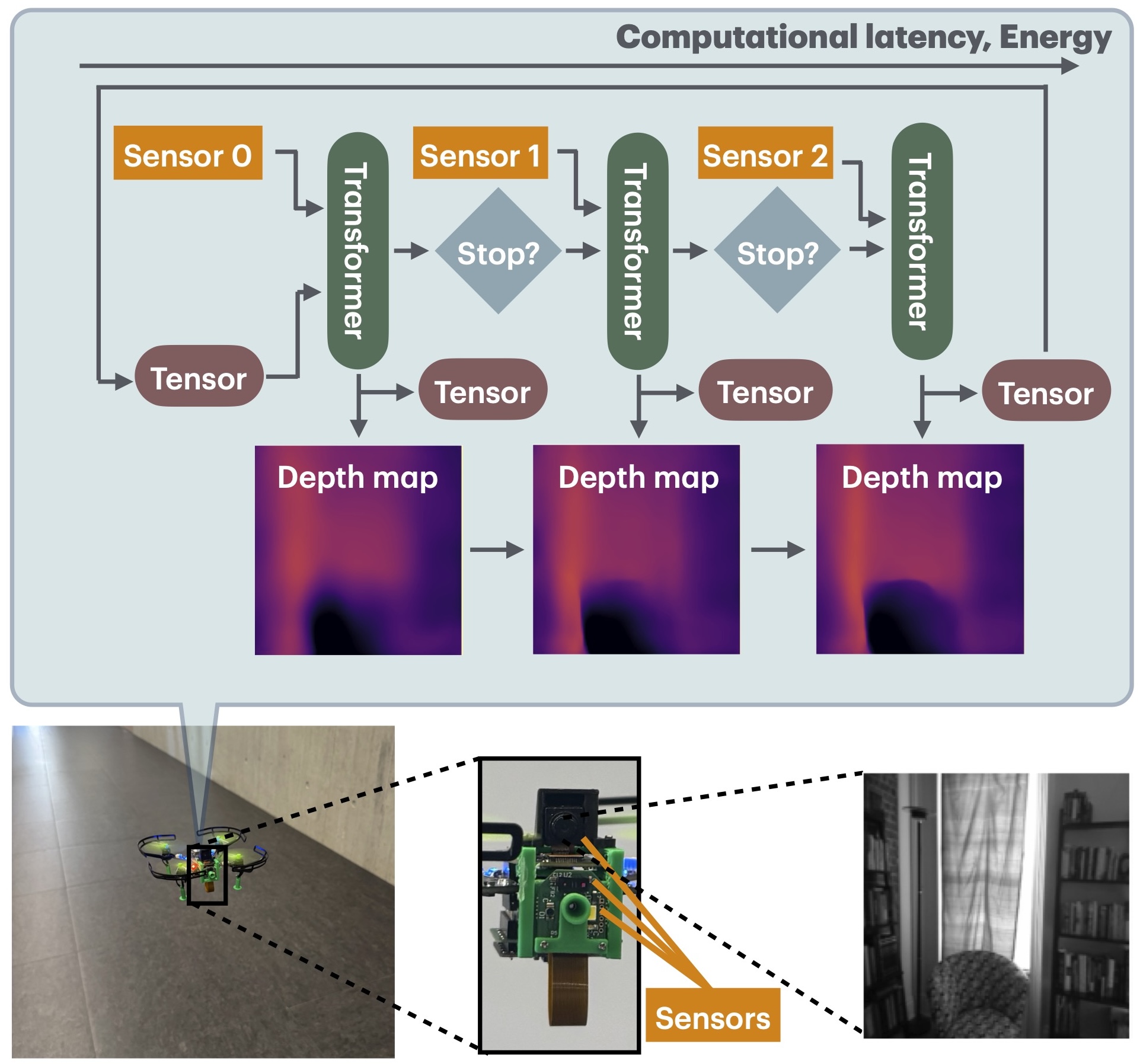}%
\end{wrapfigure}%
\begin{abstract}
Artificial intelligence (AI)-based multimodal sensor fusion is a relevant topic gaining ever more traction across ultra-low-power (ULP) embedded and cyber-physical systems, as it improves reliability, accuracy, and robustness under real-world constraints.
However, adding more and more sensors to ultra-constrained sub-\SI{100}{\milli\watt} platforms requires balancing energy consumption against prediction accuracy.
To achieve this ambitious goal, we present a novel adaptive AI methodology that combines multimodal sensor fusion (camera, ultrasound, and Time-of-Flight sensors) with a lightweight recurrent Transformer-based architecture (\SI{688}{\kilo\nothing} parameters).
We address the depth map estimation task with a mechanism that combines token propagation across iterations with incremental sensor utilization.
At each iteration, a confidence-based gating mechanism dynamically decides whether to continue the computation by adding progressively richer but more power-demanding sensors as input.
Token propagation ensures temporal consistency by forwarding context features across time.
To deploy our algorithm and test a first real-world prototype, we design a novel printed circuit board featuring all three sensors, coupled with an ULP GWT GAP9 multicore System-on-Chip.
When comparing our adaptive system against the same pipeline using all sensors and iterations on the NYUv2 dataset, we lose only 4.8\% of the $\delta_1$ accuracy in exchange for 90\% energy saving (\SI{2.44}{\milli\joule/frame}).
Finally, our adaptive method marks only 5.6\% lower $\delta_1$ accuracy than MobileDepth despite using 9$\times$ fewer parameters.
% Finally, comparing our adaptive method vs. a 9$\times$ bigger model, we lose only 5.6\% $\delta_1$ accuracy, while 
% Compared to a state-of-the-art model also running on the GAP9 but without any dynamic workload adaptation, we achieve a $\delta_1$ score 31.8\% higher while consuming the same average power ($\sim$\SI{400}{\milli\watt}).
Compared with a state-of-the-art model also running on GAP9, our method improves $\delta_1$ accuracy by 31.8\% thanks to our adaptive sensor fusion while operating within the same average power budget ($\sim$\SI{400}{\milli\watt}).
\end{abstract}

\end{minipage}}}
\maketitle

\input{1_introduction}
\input{2_related_works}
\input{3_platform}
\input{4_methods}

\input{5_results}
\input{6_conclusions}

% \appendices

% \section*{Acknowledgment}

\bibliographystyle{IEEEtran}
\bibliography{IEEEabrv,bibliography}
\newpage

\input{7_bios}

\end{document}

%% file: 1_introduction.tex
\section{Introduction} \label{sec:introduction}

Taking inspiration from biology, modern ultra-low-power (ULP) embedded systems and small-scale embodied-AI (artificial intelligence) agents are rapidly integrating novel multimodal computational methodologies to improve their accuracy, reliability, and robustness in real-world challenging scenarios~\cite{li2025ai, multimodal_smartglasses, pourjabar_sensor_fusion, kraken_shield, pulp_detector, multimodal_origami}.
However, in the context of ULP digital devices and miniaturized cyber-physical systems, designing efficient AI-based multimodal solutions remains a major challenge due to energy-accuracy trade-offs.
Achieving high prediction accuracy typically requires complex models, including State-of-the-Art (SoA) Transformer architectures~\cite{survey_edge_transformers, youtube_transformer}, while ULP platforms operate under extremely limited energy, computational, and memory budgets~\cite{pulp_detector}.
To this end, \textbf{we propose a novel, multimodal, and adaptive computational approach for ULP embodied-AI edge devices,} which we demonstrate for the dense depth map estimation task.

\begin{figure}[t]
\centering
\includegraphics[width=1.0\columnwidth]{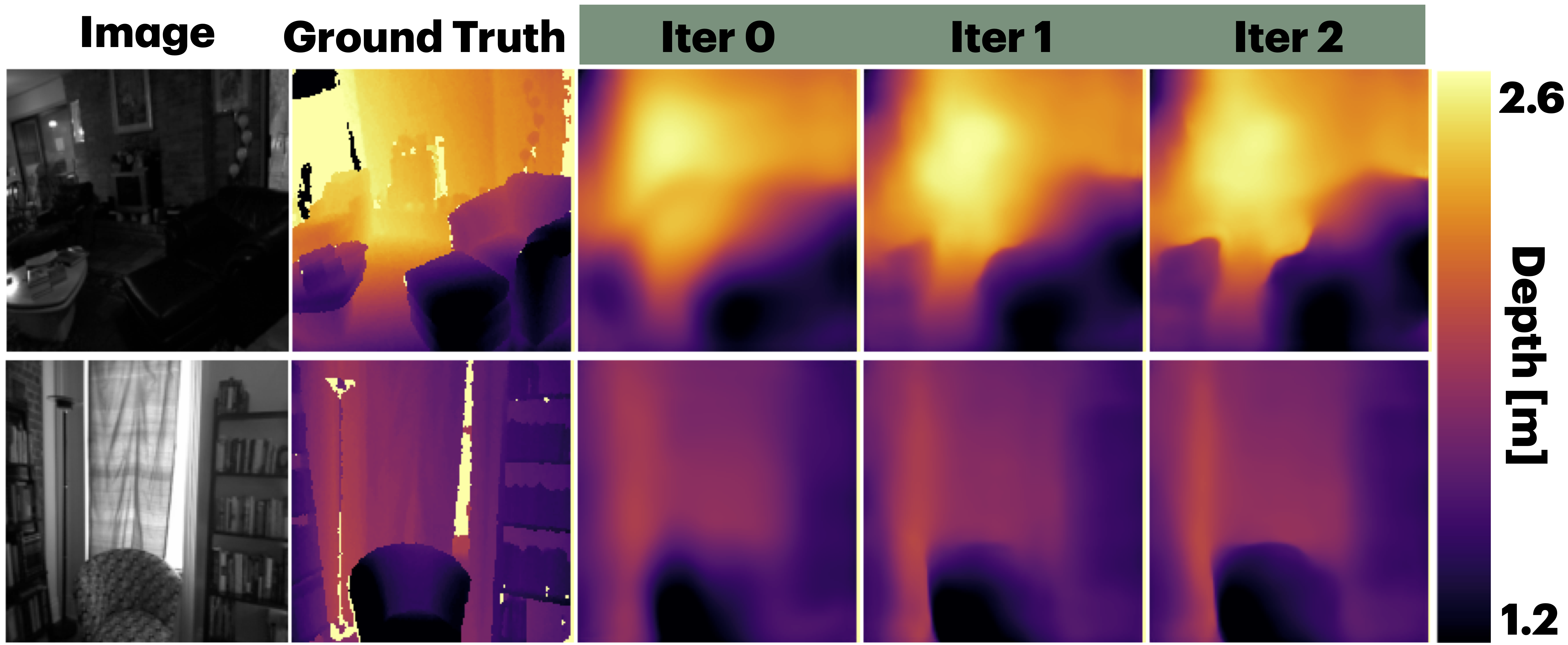}
\caption{Qualitative results on the NYUv2 dataset, our model's prediction improves iteration after iteration (i.e., iter 0, 1, and 2).}
\label{fig:depth_estimation}
\vspace{-10pt}      % shrinks the gap below this figure
\end{figure}

Depth estimation is a key capability for Internet-of-Things smart sensors monitoring/mapping environments~\cite{depth_iotdevice, depth_iotdevice2}, for wearable devices tracking users' activity~\cite{depth_est_glasses, depth_est_wearable}, and for robotic platforms performing autonomous navigation tasks~\cite{nadalini2025multi, dong2022towards}.
On the one hand, monocular depth estimation (MDE) has reached remarkable accuracy by leveraging complex vision foundation models~\cite{yang2024depth} with hundreds of millions of parameters and hundreds of billions of operations per frame, yet they exceed the capabilities of resource-constrained embedded platforms by orders of magnitude.
Depth map reconstruction directly using active depth sensors, such as Time-of-Flight (ToF) arrays or ultrasonic (US) rangers, bypasses the computational burden of vision models.
However, they can either incur significant power consumption (e.g., $\sim$\SI{300}{\milli\watt} for an 8$\times$8 ToF-based map~\cite{nadalini2025multi}) or provide very sparse resolution and a narrow field-of-view (e.g., US sensors usually collapse their sensing cone into a single scalar), or both.

To cope with this challenging scenario, we propose a novel approach that fuses three complementary sensors, i.e., a monocular camera, an US sensor, and a low-resolution ToF sensor, with a Transformer-based deep learning model~\cite{dehghani2018universal} that forwards part of its token to the next inference, providing beneficial temporal consistency.
Our approach dynamically and autonomously (with no user/external intervention) adapts the amount of sensing, computation, and power consumption based on the complexity of the input: easy scenes use a subset of sensors and minimize computation, while geometrically complex scenes rely on the more energy-intensive sensors and additional processing.
Sensorial streams are processed starting with low-resolution images and ultrasonic readings (iter 0), then high-resolution images (iter 1), and finally integrating ToF depth measurements (iter 2) as shown in Fig.~\ref{fig:depth_estimation}.
Like animals, which trade off performance for reduced brain activity to minimize cognitive energy consumption during simple tasks~\cite{kahneman2011thinking, christie2015cognitive}, our system builds upon the same approach: the \textit{``thinking fast, thinking slow''} paradigm.

Our system is designed to run on an ULP GreenWaves Technology (GWT) GAP9 multicore System-on-Chip (SoC)~\cite{pcb_gap9shield}, coupled with a Himax camera and extended with our custom printed circuit board (PCB) integrating a TDK US and an STM 8$\times$8 ToF sensors.
In summary, our key contributions are:
\begin{itemize}
    \item a novel and lightweight (\SI{688}{\kilo\nothing} parameters) Transformer-based approach for dense depth map estimation, which combines temporal token propagation across iterations with incremental multimodal sensor fusion;
    \item a confidence-based gating mechanism that dynamically decides whether to continue the computation by adding progressively richer sensorial data as input;
    % first low-resolution images and ultrasonic readings, then high-resolution images, and finally ToF depth measurements (see Fig.~\ref{fig:depth_estimation});
    \item a thorough experimental analysis of our approach, both with the standard NYUv2~\cite{silberman2012indoor} depth map dataset and with our in-field-collected dataset, which has been recorded with our custom circuit;
    \item the deployment of our system on the GAP9 SoC, including detailed latency and energy profiling.
\end{itemize}

On the NYUv2 dataset, our system scores 81.9\% $\delta_1$ accuracy when configured to always execute all three iterations with all sensors, i.e., the most energy-demanding configuration. 
Furthermore, leveraging the adaptive behavior, our algorithm achieves a $\delta_1$ of 77.1\% while reducing average energy consumption by 90\% (from 25.5 to \SI{2.44}{\milli\joule/sample}).
Notably, the confidence-based gating (\SI{4}{\kilo\nothing} additional parameters) consistently outperforms input-agnostic random early termination: at iso-energy, it reduces mean absolute error (MAE) by 29\%, while at iso-accuracy, it saves up to 60\% of energy, confirming its effectiveness in dynamically eliminating unnecessary sensing and computation.

On the GAP9 SoC, our approach achieves \SI{6}{frame/\second} with the full three-iteration pipeline and \SI{114}{frame/\second} with only the first iteration; with adaptive gating, the average throughput reaches \SI{48.3}{round/\second}, 
enabling real-time operation while dynamically adapting computation to scene complexity.
Finally, compared to a SoA model with $9\times$ more parameters~\cite{wang2020mobiledepth}, our approach loses only 5.6\% of $\delta_1$ accuracy.
Instead, by considering a SoA solution also designed for the GAP9 SoC~\cite{nadalini2025multi}, our approach improves $\delta_1$ accuracy by 27\% (77.1\% vs. 50.1\%) while being iso-power.
Ultimately, our work demonstrates that adaptive multimodal sensor fusion can effectively optimize the energy-accuracy trade-off in sub-\SI{100}{\milli\watt} ULP systems, paving the way toward the next generation of AI-based multimodal adaptive platforms.

%% file: 2_related_works.tex
\section{Related Works} \label{sec:related_works}

\begin{table}[t!]
    \centering
    \caption{Depth estimation on NYUv2 testing set.}
    \label{tab:related_work_depth_estimation_methods}
    \resizebox{\columnwidth}{!}{%
    \begin{tabular}{clccccc}
    \toprule
    & \multirow{2}{*}{\textbf{Reference}} 
    & \multirow{2}{*}{\shortstack{\textbf{Param}\\\textbf{(\SI{}{\mega\nothing})}}} 
    & \multirow{2}{*}{\shortstack{\textbf{RMSE}\\\textbf{[\SI{}{\meter}]} $\downarrow$}} 
    & \multirow{2}{*}{\shortstack{\textbf{$\delta_1$}\\\textbf{[\%]} $\uparrow$}} 
    & \multirow{2}{*}{\shortstack{\textbf{Device}}}
    & \multirow{2}{*}{\shortstack{\textbf{Power}\\\textbf{[\SI{}{\watt}]} $\downarrow$}} \\
     & & & & &   \\
    \midrule
    \multirow{6}{*}{\rotatebox{90}{\textbf{Monocular}}} & \textbf{Eigen et al.~\cite{eigen2014depth}} & 70.9 & 0.87 & 61.8 & GPU & $>$100\\ % Depth Map Prediction from a Single Image using a Multi-Scale Deep Network (Eigen et al.)
    & \textbf{Fu et al.~\cite{fu2018deep}} & 45.0 & 0.51 & 82.8 & GPU & $>$100\\ %Deep Ordinal Regression Network for Monocular Depth Estimation (Fu et al.)
    %\textbf{\cite{}} &&\\ %Pad-net: Multi-tasks guided prediction-and-distillation network for simultaneous depth estimation and scene parsing (Xu et al.)
    %& \textbf{\cite{}} & 105 & \\ % Towards Robust Monocular Depth Estimation:Mixing Datasets for Zero-shot Cross-dataset Transfer ()
    & \textbf{Ranftl et al.~\cite{ranftl2021vision}} & 343.0 & 0.36 & 90.4 & GPU & $>$100\\ %Vision transformers for dense prediction (Ranftl et al.)
    & \textbf{Yang et al.~\cite{yang2024depth}} & 335.0 & 0.21 & 98.4 & GPU &$>$100\\ %Depth Anything  (Yang et al.)
    & \textbf{Wang et al.~\cite{wang2020mobiledepth}} & 6.4 &0.50 & 82.7 & CPU & 1.8\\ %MobileDepth: Efficient Monocular Depth Prediction on Mobile Devices (Wang et al.)
    %& \textbf{} & 0.1 & & & 0.4\\ %µPyD-Net
    & \textbf{Nadalini et al.~\cite{nadalini2025multi}} & 0.1 & 0.50 & 50.1 & MCU &$<$0.4\\ % Nadalini
    \midrule
    \multirow{3}{*}{\rotatebox{90}{\textbf{Fusion}}} & \textbf{Park et al.~\cite{park2020non}} & 26.2 & 0.09 & 99.6 & GPU & $>$100\\%Non-local spatial propagation network for depth completion (Park et al.)
    & \textbf{Zhang et al.~\cite{zhang2023Completion}} & 82.6 & 0.09 & -- & GPU & $>$100\\%CompletionFormer: Depth Completion with Convolutions and Vision Transformers (Zhang et al.)
    & \textbf{Tang et al.~\cite{tang2024bilateral}} & 89.9 & 0.09 & 99.6 & GPU &$>$100\\%Bilateral Propagation Network for Depth Completion (Tang et al.)
    \midrule
    \multirow{4}{*}{\rotatebox{90}{\textbf{Temporal}}} & \textbf{Wang et al.~\cite{wang2024nvds}} & 88.3 & -- & 93.1 & GPU & $>$100\\ %NVDS+ large: Towards Efficient and Versatile Neural Stabilizer for Video Depth Estimation (Wang et al.)
    % & \textbf{\cite{luo2020consistent}} & No NYU\\ % Consistent video depth estimation (Luo et al.) 
    %& \textbf{\cite{li2023depthformer}} &  & 0.34 & 92.1 & \\%DepthFormer: Exploiting Long-Range Correlation and Local Information for Accurate Monocular Depth Estimation (Li et al.)
    & \textbf{Yang et al.~\cite{yang2024static}} & 197.0 & -- & 93.4 &  GPU & $>$100\\% STATIC : Surface Temporal Affine for TIme Consistency in Video Monocular Depth Estimation (Yang et al.)
    & \textbf{Shao et al.~\cite{shao2025learning}} & 1524.6 & -- & 94.1 & GPU & $>$100\\ %ChronoDepth: Learning Temporally Consistent Video Depth from Video Diffusion Priors (Shao et al.)
    & \textbf{Li et al.~\cite{Li_2025_CVPR}} & 1560.7 & -- & 97.3 & GPU & $>$100\\ %CH3Depth: Efficient and Flexible Depth Foundation Model with Flow Matching (Li et al.)
    \midrule
    & \textbf{Our} & 0.8 & 0.37 & 81.9 & MCU & $<$0.4\\ %All sensors included in ours
    \bottomrule
    \end{tabular}
    }
    %\vspace{-1.6em}
\end{table}

\subsection{Monocular Depth Estimation}

SoA MDE methods, summarized in Table~\ref{tab:related_work_depth_estimation_methods}, are mostly based on deep learning.
Early CNN-based approaches, such as Eigen et al.~\cite{eigen2014depth}, introduced multi-stage architectures combining coarse global predictions with local refinement, while Fu et al.~\cite{fu2018deep} reformulated MDE as a regression problem to better preserve depth discontinuities.
Ranftl et al.~\cite{ranftl2020towards} further improved generalization by training on mixed datasets and using relative depth labels for supervision.
With the advent of transformer architectures and large-scale pre-training, Ranftl et al.~\cite{ranftl2021vision} proposed dense vision transformers (ViT) that fuse multi-resolution features via CNN decoders, achieving a root mean squared error (RMSE) of \SI{0.36}{\meter} and 90.4\% $\delta_1$ accuracy on NYUv2~\cite{silberman2012indoor}.
Building on this paradigm, Yang et al.~\cite{yang2024depth} introduced Depth Anything V2, leveraging large-scale synthetic data to reach SoA performance (\SI{0.06}{\meter} MAE and 98.4\% $\delta_1$ on NYUv2).
However, these foundation models require GPU-class hardware ($>\SI{100}{\watt}$) and contain 123-\SI{335}{\mega\nothing} parameters, making them unsuitable for edge deployment.

To address resource-constrained scenarios, several lightweight MDE models have been proposed.
Wang et al.~\cite{wang2020mobiledepth} achieved real-time inference on mobile CPUs (\SI{1.8}{\watt}) using a \SI{6.4}{\mega\nothing}-parameter CNN.
More recently, Nadalini et al.~\cite{nadalini2025multi}, building on Peluso et al.~\cite{peluso2021pypydnet}, introduced a sub-\SI{100}{\kilo\nothing}-parameter model deployed on a $<\SI{400}{\milli\watt}$ MCU and fine-tuned on-device using ToF supervision, achieving \SI{0.50}{\meter} RMSE and 51.7\% $\delta_1$ on NYUv2.
However, its accuracy remains far below SoA (a 46.7\% lower $\delta_1$ accuracy than Depth Anything V2), and it requires a long and energy-hungry on-device fine-tuning phase ($\sim$\SI{860}{\second}, \SI{204.9}{\joule}), limiting its applicability to scenarios where such calibration is feasible.

Overall, as summarized in Table~\ref{tab:related_work_depth_estimation_methods}, existing methods exhibit a pronounced imbalance between model size and prediction performance. 
GPU-based foundation models such as Yang et al.\cite{yang2024depth} (\SI{335}{\mega\nothing} parameters) and Ranftl et al.\cite{ranftl2021vision} (\SI{343}{\mega\nothing} parameters) achieve up to 98.4\% $\delta_1$ accuracy, but require $>$\SI{100}{\watt} and are orders of magnitude too large for edge deployment. 
At the opposite extreme, Nadalini et al.\cite{nadalini2025multi} fits within the ULP envelope with only \SI{100}{\kilo\nothing} parameters, but at a substantial accuracy cost (50.1\% $\delta_1$). 
Wang et al.\cite{wang2020mobiledepth} occupies a middle ground with \SI{6.4}{\mega\nothing} parameters and 82.7\% $\delta_1$, but still requires \SI{1.8}{\watt}.
Our approach bridges this gap through multimodal sensing and iteration scaling rather than model-size scaling.
With only \SI{688}{\kilo\nothing} parameters our method achieves 81.9\% $\delta_1$ and \SI{0.37}{\meter} RMSE on NYUv2 while being small enough to be deployed on a sub-\SI{100}{\milli\watt} MCU. 
This represents $9\times$ fewer parameters than Wang et al., with only 0.8\% lower $\delta_1$ accuracy, over $500\times$ fewer parameters than Ranftl et al. at comparable RMSE, and a 31.8\% improvement in $\delta_1$ accuracy over Nadalini et al. with $7\times$ more parameters.
This demonstrates that combining a compact model with adaptive multimodal sensor fusion is more effective than either scaling up the model size or relying on monocular depth estimation alone.

\subsection{Multimodal Sensor Fusion for Depth Estimation}

While monocular depth estimation can achieve high accuracy in predicting relative depth~\cite{yang2024depth}, it inherently suffers from scale ambiguity.
Active ranging sensors, such as US and ToF, provide depth measurements but are typically sparse, noisy, or energy-intensive.
Multimodal depth estimation combines these complementary modalities to improve robustness and accuracy~\cite{park2020non, tang2024bilateral, zhang2023Completion}.

Early multimodal approaches focus on \emph{depth completion}, where sparse depth measurements are fused with RGB images to reconstruct dense depth maps.
Park et al.~\cite{park2020non} proposed an iterative CNN-based framework that progressively refines depth predictions using learned pixel-wise confidence propagation.
Tang et al.~\cite{tang2024bilateral} improved this paradigm through bilateral propagation, which reconstructs dense depth by propagating sparse measurements using both spatial and appearance cues.
More recently, Zhang et al.~\cite{zhang2023Completion} combined CNNs and transformers to jointly exploit the local feature extraction capabilities of convolutions and the global context modeling of self-attention.

Despite architectural differences, these methods share a common assumption: all different sensors are available and used at every inference, resulting in a fixed computational and energy cost.
From a model size perspective, multimodal fusion methods, prior to our work, required \SI{26}{\mega\nothing}--\SI{90}{\mega\nothing} parameters ($38\times$--$130\times$ larger than our model) and GPU-class hardware ($>$\SI{100}{\watt}) to achieve their accuracy. 
Our \SI{688}{\kilo\nothing}-parameter model compensates its small size by adaptively allocating sensing resources based on scene complexity, enabling energy-efficient depth prediction on a ULP GAP9 SoC at $<$\SI{0.4}{\watt}, a $250\times$ reduction in power consumption.

\subsection{Temporal Consistent Depth Estimation}

Exploiting temporal coherence in video sequences can significantly improve the accuracy of depth predictions~\cite{wang2024nvds, yang2024static, shao2025learning, Li_2025_CVPR}.
Wang et al.~\cite{wang2024nvds} propose a temporal stabilization module that, in post-processing, refines per-frame depth predictions from any monocular model by fusing past and future predictions across the video sequence.
While effective, this bidirectional design requires access to future frames, preventing its applicability to real-world scenarios.
Yang et al.~\cite{yang2024static} introduced a method that distinguishes static and dynamic regions using surface normal consistency.
Temporal stability is enhanced in static areas, while dynamic regions are aligned through feature similarity across frames.
Shao et al.~\cite{shao2025learning} proposed a diffusion-based video depth model that propagates temporal context by conditioning each inference on previous predictions, requiring multiple denoising steps per frame.
During inference, each output is initialized with previously predicted depths rather than random noise, thereby propagating temporal context across arbitrarily long sequences.
Similarly, Li et al.~\cite{Li_2025_CVPR} introduced a flow-matching depth foundation model with a latent temporal stabilizer that aggregates information from adjacent frames to enforce temporal consistency.

Despite their effectiveness, these methods have been only demonstrated on GPU-class platforms, with model sizes ranging from \SI{197}{\mega\nothing} to \SI{1560}{\mega\nothing} parameters, i.e., $250\times$ to $2000\times$ larger than our architecture, and power budgets exceeding \SI{100}{\watt}, making them unsuitable for edge deployment.
In contrast, our approach introduces \emph{temporal memory tokens} as a compact mechanism for cross-inference context propagation, adding only \SI{1}{\kilo\nothing} parameters (0.1\% overhead) while improving $\delta_1$ accuracy by 4.2\%.
Although our model achieves an 82.0\% $\delta_1$, approximately 11\%  lower than the SoA temporal depth estimation methods, it does so with up to $2000\times$ fewer parameters and allowing for deployment on MCUs. 
These results demonstrate that temporal consistency can be effectively exploited even under the stringent resource constraints of ULP embedded platforms.

\textbf{}

\subsection{Uncertainty Estimation and Adaptive Computation}

Adaptive computation offers an alternative to those approaches that require a fixed amount of computation by dynamically allocating resources based on prediction difficulty~\cite{teerapittayanon2016branchynet,huang2018multiscale,yang2025dynamicearlyexitreasoning}: inference can terminate early when predictions are confident, and continue otherwise.
Such strategies can substantially reduce average energy consumption, but they need an effective uncertainty estimation to guide resource allocation decisions.
Typically, in classification tasks, softmax outputs or learned confidence heads provide reliable stopping criteria~\cite{teerapittayanon2016branchynet, rahmath2024earlyexitsurvey}.
In contrast, dense prediction tasks such as depth estimation are more challenging, as uncertainty must be aggregated from thousands of pixel-wise predictions into a confidence signal~\cite{poggi2020uncertainty}.

Several works have explored frequency-domain cues for assessing image quality.
De et al.~\cite{de2013image} proposed an image sharpness metric based on the observation that blur suppresses high-frequency components in the Fourier domain.
More recently, Song et al.~\cite{song2025depthmaster} introduced a Fourier enhancement module within a diffusion-based monocular depth estimation model, where the U-Net hidden state is transformed via a 2D FFT and a learned modulator adaptively balances low- and high-frequency components before inverse FFT, enabling multi-frequency denoising in a single pass.
Our FFT-based confidence estimation network differs fundamentally from these approaches.
Rather than using spectral processing to enhance features or directly refine depth predictions, we leverage a lightweight neural network that processes frequency-domain similarity between the input image and the predicted depth map to estimate a scalar confidence score.
This confidence signal drives our adaptive mechanism, dynamically and automatically adding more sensors and iterations when needed.

%% file: 3_platform.tex
\section{Platform Design} \label{sec:system_design}

\begin{figure}[tb]
\centering
\includegraphics[width=1\columnwidth]{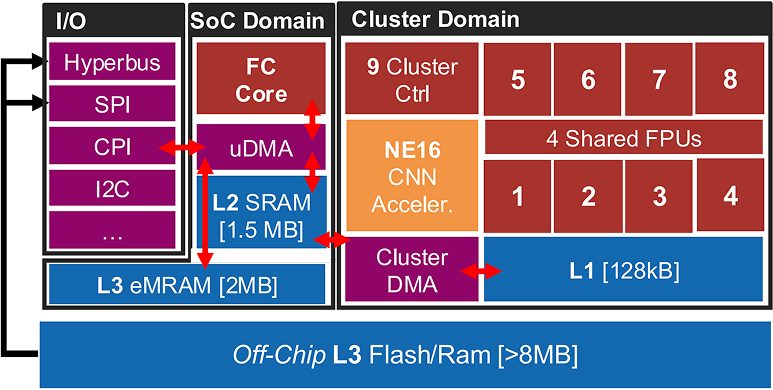}
\caption{The GAP9 SoC block diagram.}
\label{fig:gap9}
\end{figure}

\subsection{Ultra-low-power Hardware Platform} \label{subsec:hw_platform}

\begin{figure*}[t]
    \centering
    \includegraphics[width=1\linewidth]{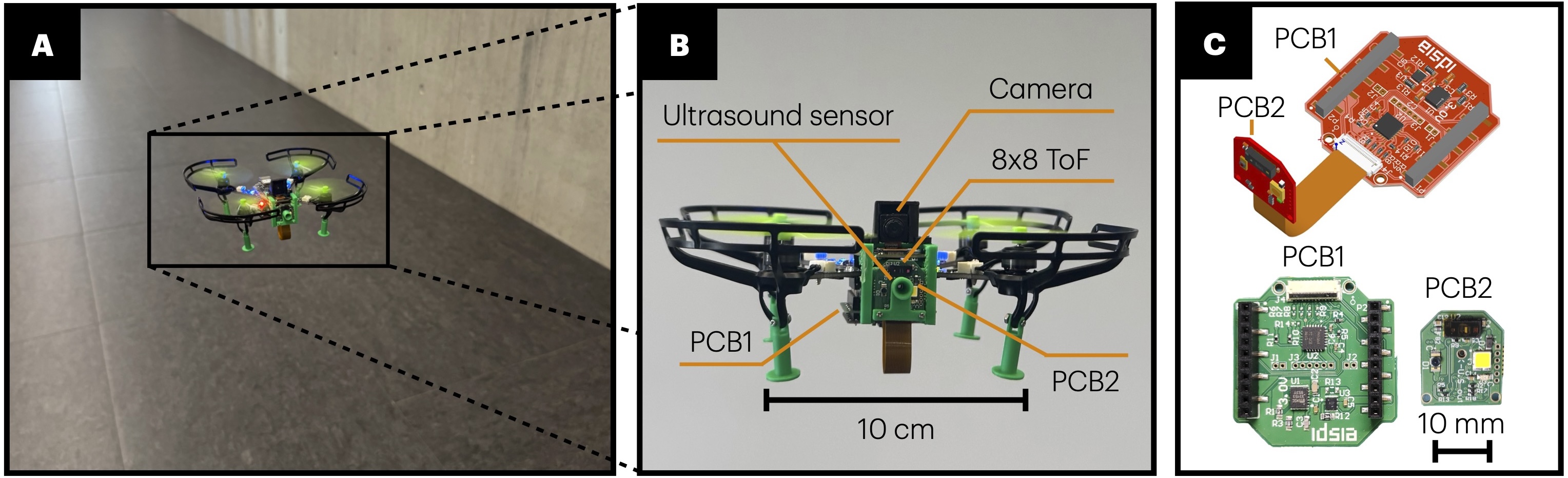}
    \caption{A) Our multimodal ranging deck deployed on a nano-drone. B) Sensors onboard. C) Renders of our multimodal ranging deck.}
    \label{fig:pcb}
\end{figure*}

The target ULP MCU used in this work is the GWT GAP9 SoC~\cite{pcb_gap9shield}, depicted in Fig.~\ref{fig:gap9}.
GAP9 features two separated processing domains: the Fabric Controller (FC) with a single RISC-V core, and the Cluster (CL) with 9 general-purpose RISC-V cores, four mixed-precision floating-point units (FPUs) (FP16/BF16/FP32), and the NE16 hardware accelerator.
NE16 supports the execution of \texttt{int8} $3\times3$ and $1\times1$ convolutions with a peak performance of \SI{150}{\giga\mac\per\second}.
All CL cores support single-instruction multiple-data (SIMD) execution, featuring a 4-lane 8-bit integer SIMD unit on each core and a 2-lane 16-bit SIMD unit on the FPUs.
The memory hierarchy of GAP9 includes \SI{128}{\kilo\byte} of shared L1 scratchpad memory and \SI{1.5}{\mega\byte} of L2 SRAM.
Accesses to L2 memory from the cluster incur an additional latency of $\sim$100 cycles.

Data transfers between L3-L2 and L2-L1 memories are handled by two direct memory accesses (DMAs), achieving a peak throughput of \SI{370}{\mega\byte/\second} and \SI{13.3}{\giga\byte/\second}, respectively.
These DMAs enable efficient overlap of computation and data movement, hiding L2 access latency for compute-bound workloads.
We use the GWT \textit{GAPflow} framework to perform post-training \texttt{int8} quantization and generate optimized C code for our transformer deployment on GAP9.
Experiments are conducted on the GAP9 evaluation kit (GAP9 EVK), which provides \SI{32}{\mega\byte} of external L3 HyperRAM.
The platform integrates a Himax HM01B0 ultra-low-power camera, supporting resolutions from $320\times320$ pixels down to QQVGA ($160\times120$ pixels), with average power consumption of \SI{2}{\milli\watt} and \SI{1.1}{\milli\watt} at \SI{30}{\fps}, respectively.

\begin{figure*}[tb]
\centering
  \includegraphics[width=\textwidth]{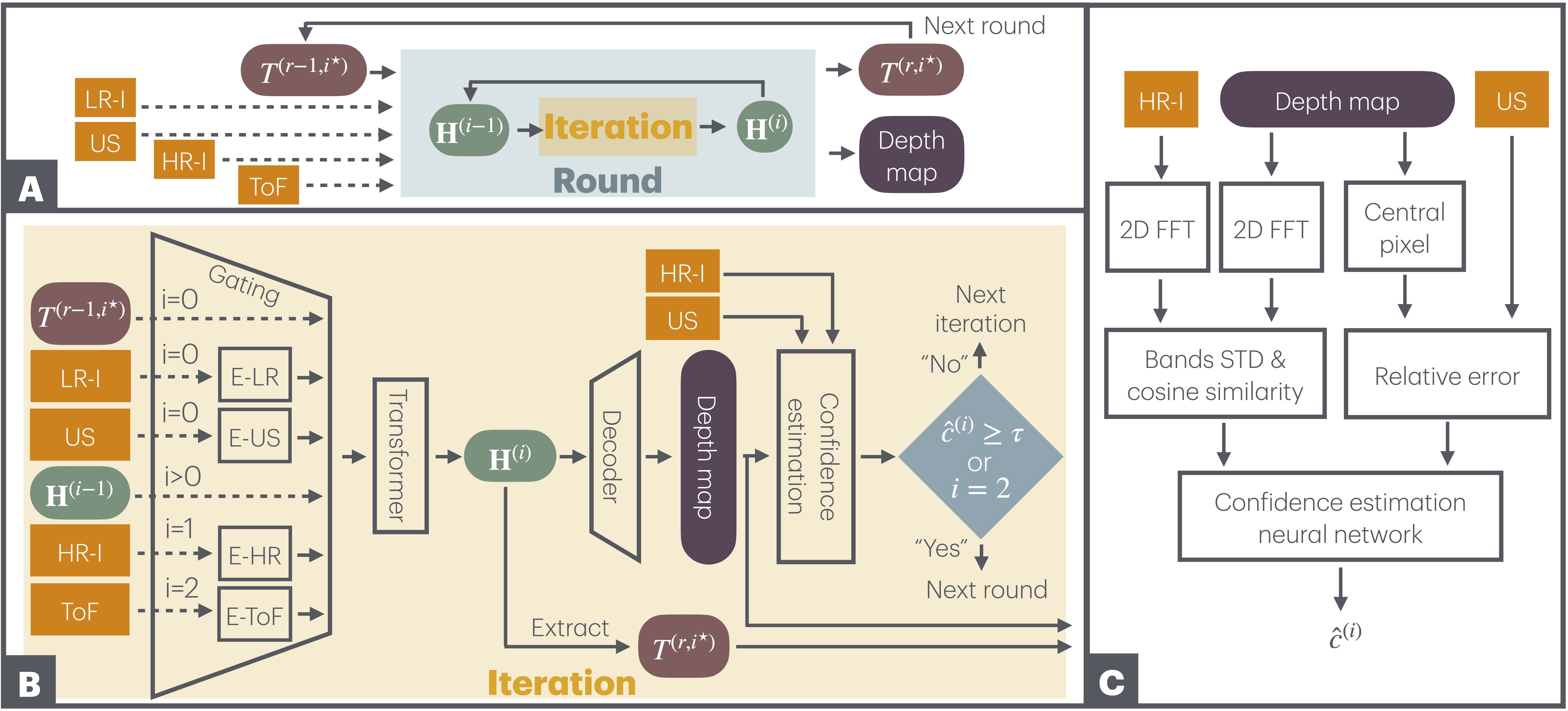}
  \caption{(A) High-level system architecture depicting how rounds (r) and iterations (i) leverage sensors and temporal data. (B) Single iteration: the transformer processes the sensorial input paired to the current iteration. Additional iterations/sensors are added depending on the prediction confidence. (C) Adaptive sensor fusion scheme.
  }
  \label{fig:transformer_architecture}
\end{figure*}

\subsection{Multimodal Active Ranging Deck} 

Our depth estimation system integrates a camera, a US sensor, and an infrared (IR) ToF sensor.
While the camera is available on the GAP9 EVK, we developed a custom PCB, the \textit{multimodal active ranging deck} (Fig.~\ref{fig:pcb}), to integrate US and ToF sensors under strict size, weight, and power constraints typical of IoT devices.
As shown in Fig.~\ref{fig:pcb}, the deck is also deployed on a nano-drone acting as a ubiquitous sensing platform, where the onboard camera complements the additional sensing modalities provided by our ranging deck.

Our deck consists of two interconnected PCBs (Fig.~\ref{fig:pcb}-C) with dimensions of 30$\times$\SI{30}{\milli\meter} and 15$\times$\SI{18}{\milli\meter}, respectively, connected via a 25-pin flexible flat cable, for a total weight of \SI{4.6}{\gram}.
The PCB1 hosts the power management and communication circuit, providing I\textsuperscript{2}C/SPI connectivity and general-purpose I/O expansion for communication with the GAP9 processor.
The power management circuit includes a switching buck regulator (MIC33153, \SI{4}{\mega\hertz} switching frequency) which allows to generate the \SI{3.0}{\volt} supply voltage from the \SI{4.2}{\volt} battery voltage with a typical efficiency of 93\%.
The PCB2 hosts two complementary active ranging sensors:

\begin{itemize}
    \item \textbf{Ultrasonic depth sensor} (TDK ICU-30201\footnote{\url{invensense.tdk.com/products/icu-30201/}}) which provides millimeter-accurate single scalar range measurements up to \SI{9}{\meter}@\SI{33}{\hertz}~\cite{pcb_batdeck}.
    The sensor features a compact package (5.17$\times$2.68$\times$\SI{0.9}{\milli\meter}) integrating a \SI{53}{\kilo\hertz} piezoelectric ultrasonic transducer and an embedded digital signal processor that preprocesses echo signals and generates single-scalar range measurements. 
    The sensor consumes \SI{1.5}{\milli\watt} during active ranging and \SI{1}{\micro\watt} when in idle mode.
    We couple the sensor with a \SI{55}{\degree} acoustic cochlea.
    \item \textbf{Multi-zone IR-based ToF sensor} (STM VL53L5CX\footnote{\url{st.com/en/imaging-and-photonics-solutions/vl53l7cx}}), which provides 8$\times$\SI{8}{\pixel} ranging measurements up to \SI{3.5}{\meter}@\SI{15}{\hertz}, with a fixed \SI{65}{\deg} diagonal field of view.
    Based on the ToF measurement principle, the sensor uses \SI{940}{\nano\meter} infrared pulses to measure and compute the absolute distances for each pixel.
\end{itemize}

Combining camera, US, and ToF sensing, as shown in Fig.~\ref{fig:pcb}-B, gives our system complementary perception capabilities.
US sensing can withstand challenging environmental conditions, such as smoke, transparent surfaces, and fog, that would affect the camera and ToF sensor.
On the other hand, the US sensor does not provide dense depth information because its sensing cone is reduced to a single measurement.
Conversely, ToF sensing provides a millimeter-scale accurate 8$\times$8 receptive field but offers a 2.5$\times$ shorter maximum distance and has the highest power consumption among our sensors, consuming approximately \SI{286}{\milli\watt}~\cite{pcb_tof} in active mode and \SI{3.3}{\micro\watt} in idle mode.

%% file: 4_methods.tex
\section{Methodology} \label{sec:method}

In this section, we describe: A) our multimodal software architecture, together with an ablation study for selecting the best backbone model among four SoA alternatives, B) our mechanism to propagate depth information through subsequent inferences, C) our confidence estimation network to enable adaptive sensor fusion, D) our training and validation procedure, and E) our pseudo-label generation procedure for recovering missing depth ground truth labels.

\subsection{Multimodal Software Architecture} \label{sec:transformer_arch}

The tight memory and power budget of ULP MCU-class processors (Section~\ref{subsec:hw_platform}) requires a compact, sub-megabyte network; we therefore identify the architecture that most accurately estimates depth under this constraint, among a multi-layer perceptron (MLP), two U-Net~\cite{ronneberger2015u} variants, two PyD-Net~\cite{poggi2018towards} variants, and a transformer.
To this end, we benchmark the candidate architectures on the NYUv2 indoor RGB-D dataset~\cite{silberman2012indoor} (see Section~\ref{subsec:dataset_training}), each fed with a single $160\times160$ grayscale image and trained to regress the corresponding dense depth map.
Table~\ref{tab:arch_comparison} compares the accuracy of all candidate models constrained to roughly the same number of parameters.
The transformer outperforms every alternative by a large margin, reaching 62.4\% $\delta_1$ accuracy and \SI{0.78}{\meter} RMSE vs. $\leq 44.5\%$ and $\geq\SI{1.18}{\meter}$ for the best-performing PyD-Net-large.
This indicates that attention-based mechanisms are more beneficial for improving accuracy rather than larger model capacity.
Motivated by this result, we adopt the transformer as the backbone of our multimodal fusion architecture, described in the remainder of this section.

\input{table/1}

To answer the key research question ``\textit{how can we enable energy-efficient sensor fusion at the extreme edge}'', we propose a recurrent transformer-based architecture for multimodal depth estimation, specifically designed for deployment on resource-constrained devices.
As illustrated in Fig.~\ref{fig:transformer_architecture}-A, the network produces dense $160\times160$ depth maps by fusing measurements from four heterogeneous and complementary sensors: low-resolution grayscale images ($40\times40$), ultrasonic scalar depth measurements, high-resolution grayscale images ($160\times160$), and low-resolution ToF depth data ($8\times8$).
The computational architecture consists of three main components: input encoders, a recurrent transformer, and output decoders.

\subsubsection{Multimodal sensor encoders} \label{subsec:encoder}

Each sensor data stream is processed by a dedicated encoder that projects the input into a shared token embedding space (Fig.~\ref{fig:transformer_architecture}-B).
We label the encoders for low- and high-resolution images as E-LR and E-HR, respectively, and those for ultrasonic and ToF measurements as E-US and E-ToF.
Following the patch-based embedding strategy commonly adopted in Vision Transformers~\cite{dosovitskiy2020image}, the image-based inputs and the ToF depth data are partitioned into non-overlapping $8\times8$ patches.
Specifically, the $40\times40$ low-resolution image, the $160\times160$ high-resolution image, and the $8\times8$ ToF are transformed into $5\times5$, $20\times20$, and $1\times1$ token grids, respectively.
Each set of input patches is projected into a sequence of tokens using a single convolutional layer with kernel size and stride equal to the patch size, yielding 25, 400, and 1 tokens, respectively, each with embedding dimension $D=128$.
The ultrasonic depth measurement is encoded through a linear projection, producing a single token with the same embedding dimension $D=128$.
For all data streams, positional embeddings are added to the tokens to encode the temporal order in which sensors are fed to the transformer block across recurrent iterations.

\begin{figure*}[tb]
\centering
\includegraphics[width=1.0\linewidth]{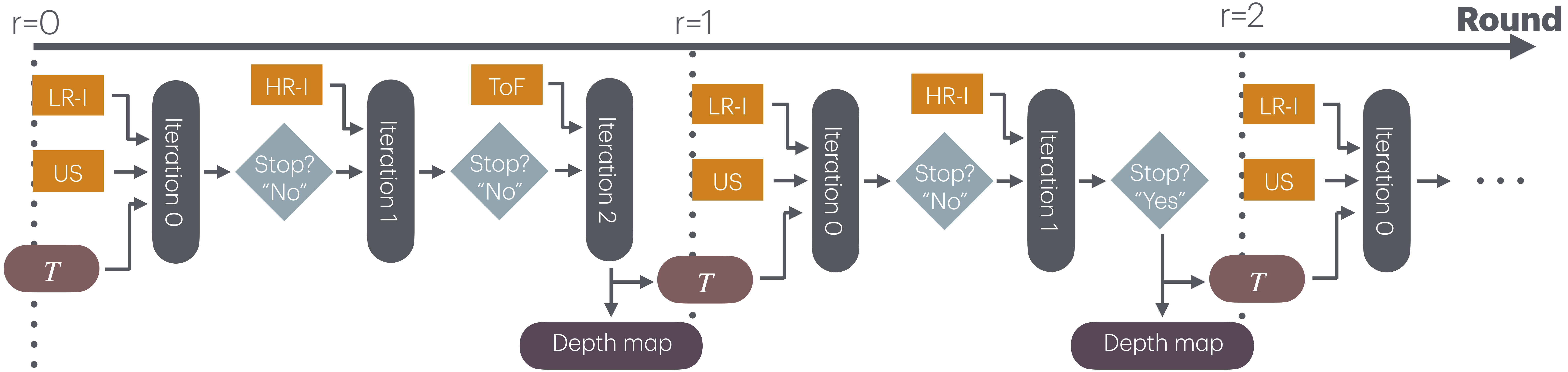}
\caption{\textit{Rounds} of our sensor fusion neural network. Inputs are: Image Low Resolution (I-LR), Ultrasound Sensor (US), Image High Resolution (I-HR), Time-of-Flight (ToF). Temporal memory tokens ($\mathbf{T}$) range from 0 to 64. At the round r=0, the $\mathbf{T}$ are initialized randomly.}
\label{fig:timeline_sensor_fusion}
\end{figure*}

\subsubsection{Recurrent transformer} \label{subsec:transformer_arch}

inspired by the Universal Transformer~\cite{dehghani2018universal}, our architecture employs a single recurrent transformer block that iteratively refines the output tokens through the output-to-input feedback, as shown in Fig.~\ref{fig:transformer_architecture}-B.
The overall execution flow of the proposed sensor-fusion mechanism is illustrated in Fig.~\ref{fig:timeline_sensor_fusion}.
An \emph{iteration} $i$ corresponds to a single inference of the transformer: it takes as input the current sensor tokens ($\mathbf{X}^{(i)}_{sensors}$) together with the output tokens from the previous iteration $\mathbf{H}^{(i-1)}$, and produces updated tokens $\mathbf{H}^{(i)}$, which are decoded into a depth map prediction (Section~\ref{subsec:decoder}).
A \emph{round} denotes one complete depth-estimation process and consists of one or more iterations that progressively refine the prediction until a confidence criterion is satisfied (Section~\ref{subsec:confidence_estimation}).
Recurrence is confined within a round: the output tokens of iteration $i$ are fed as input to iteration $i\!+\!1$, while no token information is propagated across \textit{rounds.}

This design is particularly well-suited to resource-constrained devices.
Reusing the same transformer block across iterations enables parameter sharing and a compact model size, while progressive sensor integration supports adaptive sensor fusion, allowing low-power sensor data to be processed first and more energy-expensive sensor data to be incorporated only when additional refinement is required.
Importantly, this incremental design also provides inherent robustness to partially missing or degraded sensor data: since the architecture is trained to produce valid depth predictions at every iteration using only the sensors available up to that point, the system naturally operates with any subset of its sensor suite, from a single low-resolution image and ultrasound reading at iteration 0, to the full multimodal set at iteration 2.
We call this execution scheme \emph{progressive sensor fusion} (\textbf{fixed iter}) and, in our setup, we do up to three iterations per round.
The first iteration (\textit{i}=0) processes the US sensor data and a low-resolution (40$\times$\SI{40}{\pixel}) image (I-LR), with a total processing cost of \SI{0.49}{\milli \joule} including both sensor data.
The second iteration (\textit{i}=1) processes a high-resolution (160$\times$\SI{160}{\pixel}) image (I-HR) with a \SI{2.02}{\milli\joule/sample} energy requirement, while the third iteration (\textit{i}=2) incorporates the ToF sensor data (\SI{2.02}{\milli\joule/sample}).
Ultimately, the transformer input at round $r$ and iteration $i$ is defined as:
\begin{equation*}
\mathbf{X}_{\text{in}}^{(r,i)} =
\begin{cases}
\mathbf{X}_{\text{sensors}}^{(r,0)},
& i = 0, \\[8pt]

\big[\, 
\mathbf{X}_{\text{sensors}}^{(r,i)} \; ; \;
\mathbf{H}^{(r,i-1)}
\,\big],
& i \in \{1,2\}.
\end{cases}
\end{equation*}
where $[\cdot \, ; \, \cdot]$ denotes tokens' concatenation.

More generally, our transformer architecture is modular and sensor-agnostic: new sensor streams can be integrated by adding dedicated encoders that project measurements into the shared token space, without altering the recurrent transformer itself.

\subsubsection{Decoder} \label{subsec:decoder}
At each iteration, the transformer outputs a set of tokens corresponding to all processed sensor streams.
Only tokens originating from the image encoders are decoded into a dense $1\times160\times160$ depth map $\mathbf{D}^{(i)}$, while tokens from non-image sensors are used exclusively to refine the latent representations through attention.
We employ two distinct lightweight decoders, selected according to the spatial resolution of the available image tokens.
When only low-resolution image tokens are present ($i=0$), a low-resolution decoder operates on a coarse $5\times5$ token grid and progressively upsamples it to full resolution through five successive stages, each doubling the spatial dimensions.
Once high-resolution image tokens are introduced ($i=1,2$), a high-resolution decoder is used; it operates on a finer $20\times20$ token grid and produces the final $160\times160$ depth map via three $2\times$ upsampling stages.
Each upsampling stage consists of a $3\times3$ convolution that halves the channel dimension, followed by group normalization, ReLU activation, and $2\times$ bilinear interpolation. 
The final prediction head applies a $3\times3$ convolution reducing to 16 channels, followed by a $1\times1$ convolution to produce the single-channel depth output, normalized to $[0,1]$ via sigmoid activation. 

\textbf{Model size.}
The proposed architecture is designed to operate within the tight memory constraints of resource-constrained edge platforms.
The complete depth estimation network comprises a total of \SI{683}{\kilo\nothing} trainable parameters.
The four sensorial encoders account for \SI{135}{\kilo\nothing} parameters.
In particular, the encoder for high-resolution images contains \SI{162}{\kilo\nothing} parameters, while the low-resolution image encoder uses \SI{18}{\kilo\nothing} parameters.
The encoder for $8\times8$ ToF depth maps contributes \SI{9}{\kilo\nothing} parameters, and the ultrasonic depth encoder requires only \SI{0.8}{\kilo\nothing} parameters.

The transformer comprises \SI{259}{\kilo\nothing} parameters, split between learnable positional embeddings (\SI{60}{\kilo\nothing}) and a single shared transformer block (\SI{199}{\kilo\nothing}).
Finally, the depth decoders account for the remaining parameters: \SI{116}{\kilo\nothing} for the high-resolution decoder and \SI{118}{\kilo\nothing} for the low-resolution decoder.
Overall, the compact parameter footprint allows the \texttt{int8}-quantized model (\SI{683}{\kilo\byte}) to fit entirely within the on-chip L2 SRAM of the GAP9 platform, occupying less than 46\% of the available \SI{1.5}{\mega\byte} and leaving the remaining memory available for runtime activation buffers, thereby satisfying the stringent memory constraints of ultra-low-power edge deployment without requiring external L3 memory access for model weights.

\subsection{Temporal Memory Tokens} \label{subsec:temporal_tokens}

Although the architecture described in Section~\ref{sec:transformer_arch} does not propagate information across successive rounds, intelligent sensor nodes deployed in real-world scenarios operate on continuous and temporally correlated sensor streams, where consecutive frames share significant geometric structure (e.g., static surfaces, slowly moving objects).
To exploit this temporal coherence, we extend the architecture with a lightweight form of temporal memory that carries depth-relevant context from one round to the next, enabling the transformer to reuse previously learned scene representations rather than estimating each frame from scratch.
Specifically, we introduce an additional set of $N_t$ input tokens, referred to as \emph{temporal memory tokens} $\mathbf{T}$, as shown in Fig.~\ref{fig:transformer_architecture}.
These tokens are randomly initialized once before any inference (round 0), and provided as input to the transformer at every iteration.
Within a round, temporal memory tokens are progressively refined through the recurrent transformer across multiple iterations, absorbing information from each sensor modality via bidirectional self-attention;
their final state is then propagated to the next \textit{round}, where they provide the transformer with a prior on the scene's depth structure before any new sensor data is processed.
This propagation mechanism improves temporal consistency in two ways: \textit{(i)} it biases the current prediction toward the geometric structure observed in previous frames, reducing frame-to-frame depth fluctuations in static regions, and \textit{(ii)} it provides a warm-start that allows the model to reach accurate predictions with fewer iterations, as the temporal context partially compensates for sensor modalities that have not yet been processed in the current round.
Ultimately, for round $r$ and iteration $i$, the transformer input is defined as:
\begin{equation*}
\mathbf{X}_{\text{in}}^{(r,i)} =
\begin{cases}
\big[\,
\mathbf{X}_{\text{sensors}}^{(r,0)} \; ; \;
\mathbf{T}^{(r-1,i^\star)}
\,\big],
& i = 0, \\[6pt]

\big[\,
\mathbf{X}_{\text{sensors}}^{(r,i)} \; ; \;
\mathbf{H}^{(r,i-1)}
\,\big],
& i \in \{1,2\}.
\end{cases}
\end{equation*}
where $i^{\star}$ is the last iteration of round $r-1$, and $\mathbf{T}^{(r,i-1}) \in \mathbf{H}^{(r,i-1})$

As discussed in Section~\ref{sec:transformer_arch}, the hidden-state tokens $\mathbf{H}$ are reset at the beginning of each \textit{round}, as propagating the full hidden state across \textit{rounds} would lead to an unbounded and exponential growth in the number of tokens, making inference computationally infeasible.
In contrast, we only propagate temporal memory tokens across rounds, which introduce a constant overhead of $N_t$ tokens.
We evaluate configurations varying  $N_t \in [0, 16]$.
Each temporal memory token has an embedding dimension $D=128$, and adds $\sim$\SI{0.5}{\kilo\nothing} learnable parameters, representing less than 0.1\% overhead compared to the rest of the model.

\subsection{Confidence Estimation Network}
\label{subsec:confidence_estimation}

To allow our transformer to dynamically and autonomously adapt the amount of sensing, computation, and power consumption required for each depth prediction, we introduce an adaptive sensor fusion mechanism that explicitly links scene complexity to the number of iterations executed.
Specifically, we employ a lightweight confidence estimation network, which evaluates the predicted depth map at each iteration and decides whether to terminate inference or continue refining by fusing additional sensors. 
Geometrically simple scenes, such as flat walls, empty corridors, or uniform surfaces, produce high-confidence predictions after the first iteration using only low-cost sensors and terminate early. 
Geometrically complex scenes, containing cluttered objects, fine-grained depth discontinuities, or textured surfaces at varying depths, yield low-confidence predictions that trigger additional iterations, progressively incorporating the high-resolution camera and the ToF sensor to resolve ambiguities.

As shown in Fig.~\ref{fig:transformer_architecture}-C, scene complexity is quantified through the frequency-domain structure of the input image and the predicted depth map $\hat{\mathbf{D}}^{(i)}$.
The key intuition is that reliable depth predictions should exhibit frequency characteristics similar to those of the input image: image edges typically correspond to depth discontinuities, while smooth image regions correspond to planar surfaces.
Therefore, a strong spectral similarity indicates that the scene geometry has been adequately captured, allowing early termination, while a spectral mismatch signals unresolved geometric complexity and triggers additional sensor fusion.

At each iteration $i$, we compute the 2D Fast Fourier Transform (FFT) of the input grayscale image and the predicted depth map, and extract their magnitude spectra.
Frequencies are normalized to $[0,1]$ and divided into ten equally spaced bands in $[0,0.5]$, where $0.5$ corresponds to the Nyquist frequency.
For each band, we compute two complementary features:
\textit{i}) the \textbf{cosine similarity} $\in [0,1]$ between the image and depth magnitude spectra, where a ratio close to 1 indicates a similar distribution in the frequency content, and
\textit{ii}) the \textbf{ratio of their standard deviations}, capturing whether the depth prediction exhibits variability comparable to the image.

\begin{figure}[t]
\centering
\includegraphics[width=1.0\columnwidth]{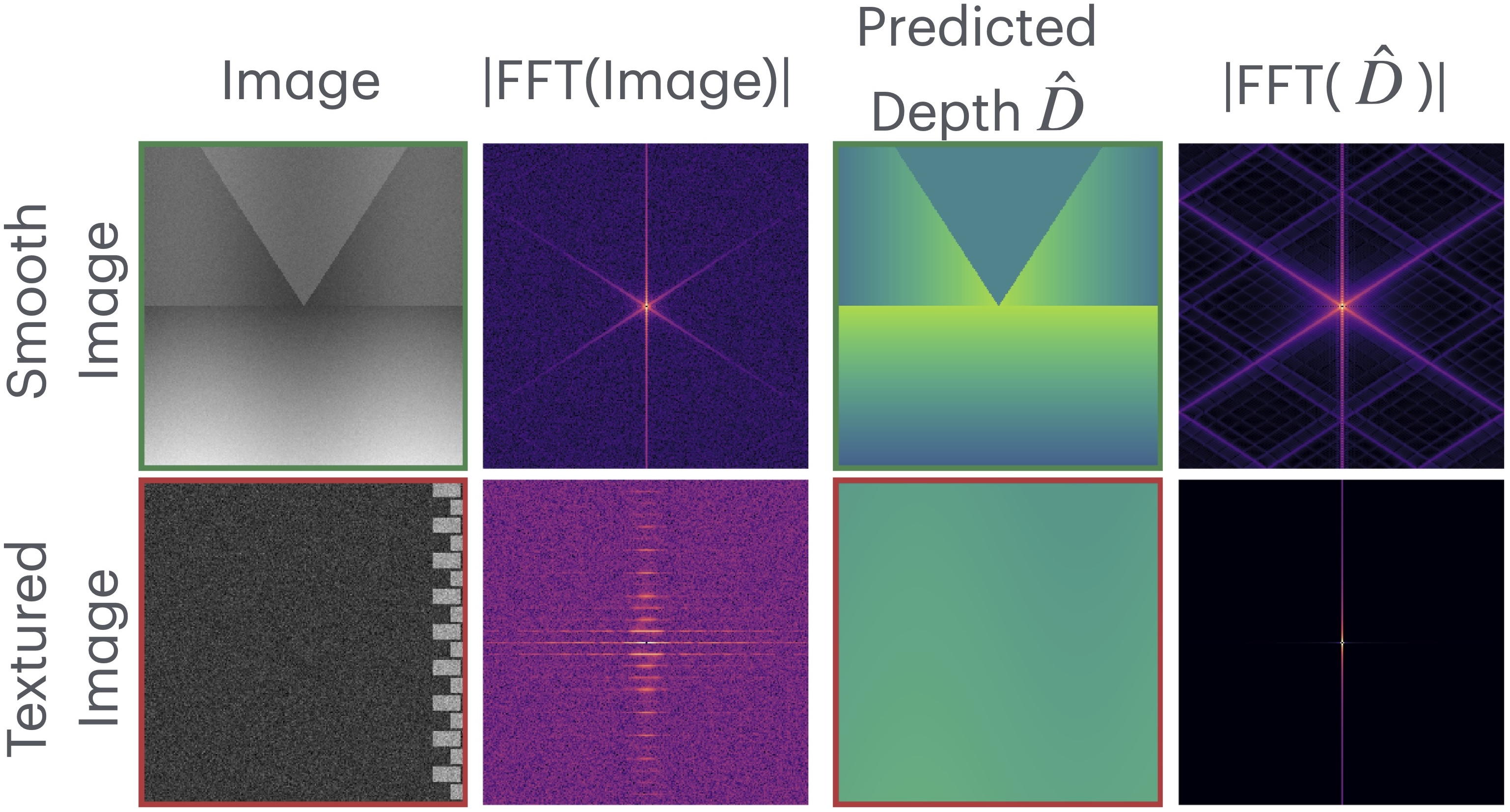}
\caption{
Comparison of the FFT of two sample grayscale images (smooth vs. textured) and the FFT of two respective predicted depths. The smooth sample yields high confidence in the two FFTs and early termination, whereas the textured sample yields the opposite.
}
\label{fig:fft_toy_example}
\end{figure}

Together, these features quantify how well the predicted depth reproduces the image's frequency content.
This analysis naturally captures cases where visual appearance alone is insufficient to resolve the underlying geometry, as illustrated by the two examples in Fig.~\ref{fig:fft_toy_example}.
For instance, highly textured planar surfaces introduce high-frequency components in the image but not in the depth map; the resulting spectral mismatch yields low confidence, triggering additional iterations that incorporate the ToF sensor to resolve the depth ambiguity through direct measurement.
Conversely, a scene with clearly defined geometric structure (e.g., a corridor with distinct walls and floor) produces high spectral similarity after the first iteration, enabling early termination with only the low-cost camera and ultrasound sensors.
Furthermore, the analysis naturally captures ambiguous cases and is robust to noisy sensor conditions: when a sensor produces unreliable measurements, the resulting depth prediction will exhibit spectral inconsistencies with the input image, yielding low confidence and triggering additional iterations that incorporate complementary sensors to compensate for the degraded modality.

The ten frequency bands yield 20 FFT-based features.
In addition, we compute the relative error as $\frac{|x_{depth}- x_{us}|}{x_{us}}$, where  $x_{depth}$ is the center pixel of the predicted depth map, and $x_{us}$ is the scalar US measurement.
All features are processed by a lightweight MLP followed by a sigmoid activation, producing a confidence score $\hat{c}^{(i)} \in [0,1]$.
The round terminates when $\hat{c}^{(i)}$ exceeds a threshold $\tau$.
The confidence estimation network adds only \SI{4}{\kilo\nothing} trainable parameters, resulting in a total model size of \SI{687}{\kilo\nothing} parameters.

\begin{figure}[t]
\centering
\includegraphics[width=\columnwidth]{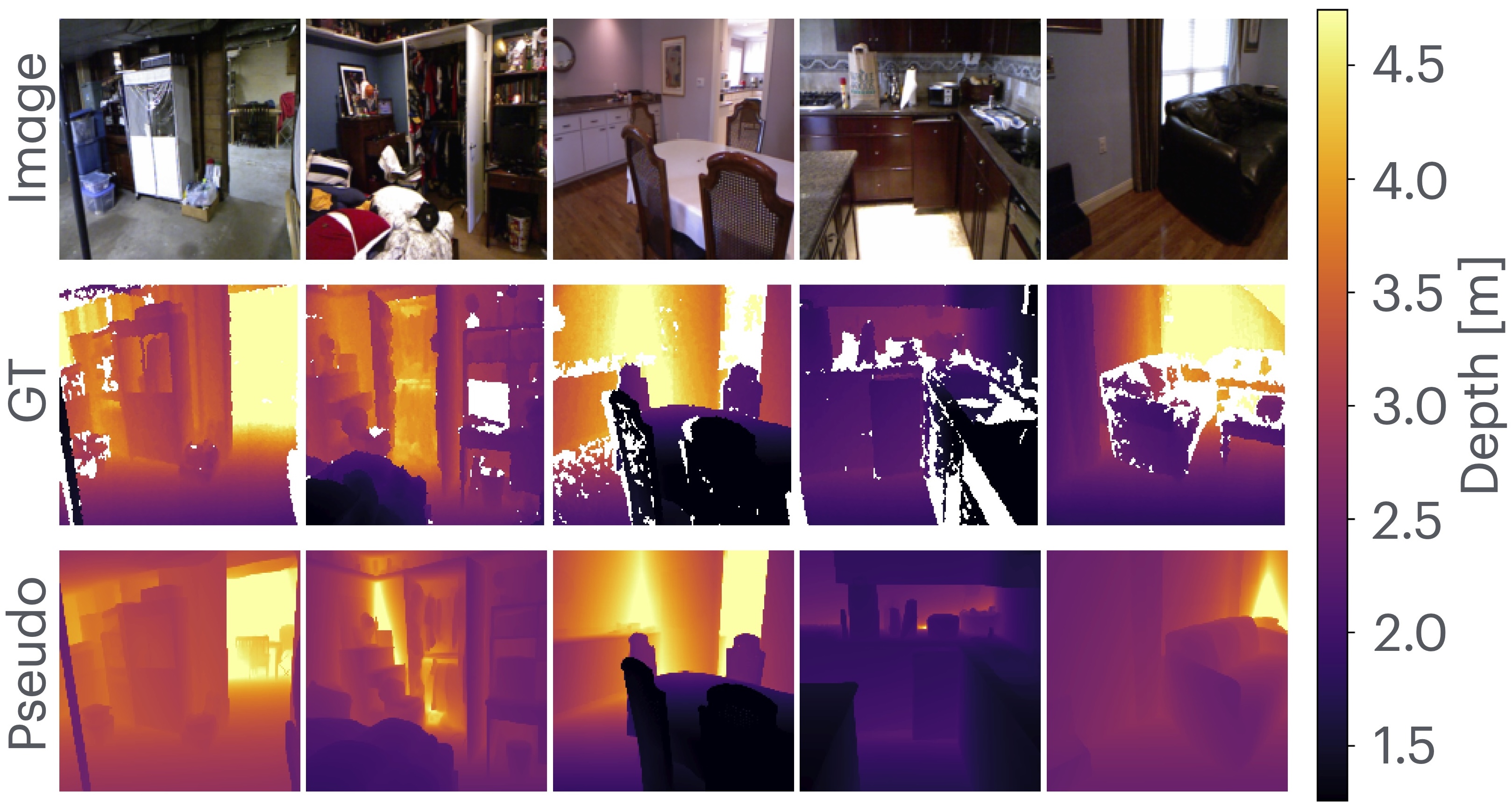}
\caption{Comparison of images, Kinect GT depths, and the P-GT labels we generated with Depth Anything V2~\cite{yang2024depth} on NYUv2 training samples.}
\label{fig:depth_comparison}
\end{figure}

\subsection{Dataset and Model Training}
\label{subsec:dataset_training}

\textbf{Dataset.} We train and evaluate our model on the NYUv2  dataset~\cite{silberman2012indoor}, which consists of RGB-D video sequences captured across 464 diverse indoor scenes, spanning bedrooms, kitchens, offices, living rooms, bathrooms, and other room types, using a Microsoft Kinect sensor, providing a broad testbed for assessing robustness across varying layouts, furniture arrangements, lighting conditions, and depth ranges.
The dataset is divided into \SI{339.4}{\kilo\nothing} training samples, \SI{97.9}{\kilo\nothing} validation samples, and \SI{69.3}{\kilo\nothing} test samples.
Each sample comprises  $640\times480$ RGB image and its corresponding depth ground truth (GT).
To emulate the sensing setup of our hardware platform (Section~\ref{subsec:hw_platform}), we pre-process the data as follows:
\textit{i}) RGB images are converted to grayscale and resized to $160\times160$ pixels to match the resolution of the Himax camera;
\textit{ii}) ToF measurements are emulated by applying min-pooling over the ground truth depth map to obtain an $8\times8$ low-resolution depth grid;
\textit{iii}) US measurements are simulated by extracting the depth value at the image center from the ground truth depth map.
All depth values are normalized to the range $[0,1]$, where a value of $1$ corresponds to a depth of \SI{10}{\meter}.
Finally, we adopt two different training procedures depending on whether temporal memory tokens are employed, as described below.

\textbf{Training without temporal memory tokens.}
The model is trained in two stages using the PyTorch framework.
In the first stage, the sensor encoders, recurrent transformer, and decoders are trained jointly for 200 epochs with a batch size of 32, using the AdamW optimizer ($\eta=10^{-4}$, weight decay $10^{-4}$) and a cosine annealing learning-rate schedule.
For each \textit{round}, we compute the loss function as the sum of BerHu~\cite{zwald2012berhu} loss over the three iterations:
\begin{equation}
\mathcal{L} = 
    \sum_{i=0}^{2} 
    \mathcal{L}_{\text{BerHu}}(\hat{\mathbf{D}}^{(i)}, \mathbf{D}_{\text{GT}})
\end{equation}
where $\hat{\mathbf{D}}^{(i)}$ is the depth map prediction at iteration $i$, and $\mathbf{D}_{GT}$ is the ground truth depth.

Training early stopping with a patience of 15 epochs is applied based on the validation loss.
In the second stage, the depth estimation network is frozen, and only the confidence estimation networks are trained for 50 epochs using the same loss function, a reduced learning rate of $\eta=5\times10^{-5}$, and a batch size of 32.

\textbf{Training with temporal memory tokens.}
We follow the same two-stage training procedure, modifying only the first stage.
Specifically, we sample contiguous sequences of $F=16$ frames from NYUv2 video trajectories and process them sequentially while propagating temporal memory tokens across frames.
The cumulative training loss $\mathcal{L}_{\text{temporal}}$ aggregates the BerHu loss over both frames and iterations as follows:
\begin{equation}
\mathcal{L}_{\text{temporal}} = 
\frac{1}{F} 
\sum_{f=0}^{F-1} 
    \sum_{i=0}^{2} 
        \mathcal{L}_{\text{BerHu}}(\hat{\mathbf{D}}^{(f,i)}, \mathbf{D}_{\text{GT}}^{(f)})
\end{equation}
where $\hat{\mathbf{D}}^{(f,i)}$ is the depth map predicted at frame $f$ and iteration $i$, and $\mathbf{D}^{(f)}_{GT}$ is the ground truth depth associated with frame $f$.
To ensure that temporal memory tokens remain informative under adaptive sensor fusion, we terminate inference rounds at random iterations during training.
Temporal memory tokens are reset at the end of each frame sequence to prevent information leakage between unrelated scenes.

\input{table/2}

\begin{figure}[t]
\centering
\includegraphics[width=\columnwidth]{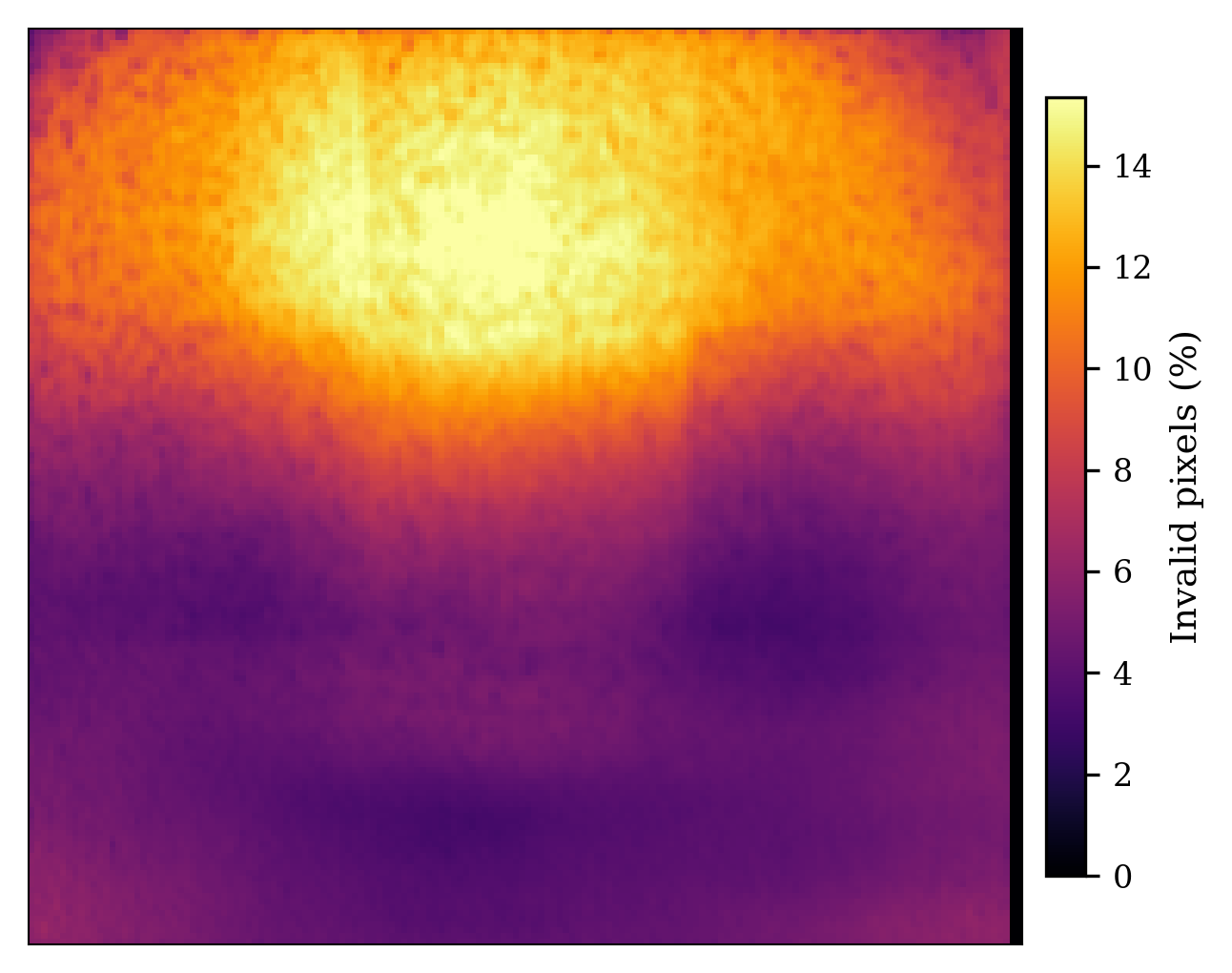}
\caption{Heatmap representing spatial distribution of invalid Kinect depth measurements across the NYUv2 training set. Brighter regions indicate a higher percentage of pixels with missing GT depth label.}
\label{fig:kinect_invalid_heatmap}
\end{figure}

\subsection{Pseudo-label Generation}
\label{subsec:pseudo_labels}
The NYUv2 dataset provides depth GT labels acquired with a Microsoft Kinect sensor, which suffer from noise and missing measurements, particularly at object boundaries, reflective surfaces, and regions beyond the sensor’s operating range (Fig.~\ref{fig:depth_comparison}).
Fig.~\ref{fig:kinect_invalid_heatmap} shows the spatial distribution of invalid Kinect depth measurements across the NYUv2 training set.
On average, 7.8\% of pixels lack valid depth, with invalid rates reaching up to $\sim$16\% in the upper regions of the image.
Such sparse and noisy supervision can hinder training by biasing the model toward learning sensor-specific artifacts rather than geometrically consistent depth cues.

To mitigate this issue, we generate new dense pseudo-labels (P-GT) for the training set using Depth Anything V2~\cite{yang2024depth}, a SoA monocular depth estimation foundation model.
We employ its most accurate variant, Depth Anything V2-Large (\SI{335}{\mega\nothing} parameters), to produce artifact-free depth predictions for all training images.
As Depth Anything V2 outputs relative depths, we align the pseudo-labels to the metric scale of the Kinect ground truth by matching the 10th and 90th depth percentiles over valid pixels via a linear transformation.
The test set is instead left unchanged the original Kinect GT labels.
Fig.~\ref{fig:depth_comparison} qualitatively compares Kinect ground truth and Depth Anything V2 pseudo-labels on representative NYUv2 samples, highlighting that the pseudo-labels are dense and free of visible artifacts.

%% file: table/1.tex
\begin{table}[t!]
    \centering
    \caption{Backbone selection on the NYUv2 dataset.}
    \label{tab:arch_comparison}
    \resizebox{\columnwidth}{!}{%
\begin{tabular}{lccc}
\toprule
\textbf{Model} &
\textbf{$\delta_1$ $\uparrow$ [\%]} &
\textbf{RMSE $\downarrow$ [\si{\meter}]} &
\textbf{\# params [\si{\mega\nothing}]} \\
\midrule
MLP & 41.8 & 1.26 & 0.65 \\
% \addlinespace
U-Net-large~\cite{ronneberger2015u} & 43.5 & 1.22 & 1.02 \\
U-Net-small~\cite{ronneberger2015u} & 43.6 & 1.21 & 0.45 \\
% \addlinespace
PyD-Net-large~\cite{poggi2018towards} & 44.5 & 1.18 & 0.91 \\
PyD-Net-small~\cite{poggi2018towards} & 44.2 & 1.20 & 0.41 \\
% \addlinespace
\textbf{Transformer (ours)} & \textbf{62.4} & \textbf{0.78} & 0.68 \\
\bottomrule
\end{tabular}%
}
\end{table}

%% file: table/2.tex
\begin{table*}[t!]
    \centering
    \caption{NYUv2 testing set results always halting the computation at the specified iteration. Ground Truth (GT), Pseudo-labels (P-GT). Inputs are: Image Low Resolution (I-LR), Ultrasound Sensor (US), Image High Resolution (I-HR), Time-of-Flight (ToF).}
    \label{tab:nyuv2_energy_results}
    \resizebox{\textwidth}{!}{%
\begin{tabular}{ccccccccccccc} 
\toprule
\multirow{2}{*}{\shortstack{\textbf{Training} \\ \textbf{labels}}} & 
\multirow{2}{*}{\textbf{Configuration}} & 
\multirow{2}{*}{\textbf{Iteration}} & 
\multicolumn{4}{c}{\textbf{Inputs}} & 
\multirow{2}{*}{\textbf{MAE $\downarrow$ [\SI{}{\meter}]}} & 
\multirow{2}{*}{\textbf{RMSE $\downarrow$ [\SI{}{\meter}]}} & 
\multirow{2}{*}{\textbf{$\delta_1$ $\uparrow$ [\%]}} & 
\multirow{2}{*}{\textbf{$\delta_2$ $\uparrow$ [\%]}} & 
\multirow{2}{*}{\textbf{$\delta_3$ $\uparrow$ [\%]}} & 
\multirow{2}{*}{\shortstack{\textbf{Energy/round} $\downarrow $\\ \relax [\SI{}{\milli\joule\per\round}]}} \\ 
\cmidrule(l){4-7}
 &  &  & I-LR & US & I-HR & ToF &  &  &  &  &  &  \\ 
\midrule
\multirow{5}{*}[-5pt]{GT} & \multirow{3}{*}{\textbf{Fixed iter}} & \textbf{Iter 0} & \checkmark & \checkmark & $\times$ & $\times$ & 0.58 & 0.79 & 55.9 & 74.0 & 82.9 & 0.5 \\
 &  & \textbf{Iter 1} & $\times$ & $\times$ & \checkmark & $\times$ & 0.56 & 0.76 & 59.0 & 75.9 & 84.5 & 2.5 \\
 &  & \textbf{Iter 2} & $\times$ & $\times$ & $\times$ & \checkmark & 0.32 & 0.51 & 75.6 & 86.3 & 90.7 & 25.4 \\ 
\addlinespace
 & \textbf{All-always} & \textbf{Iter 0} & \checkmark & \checkmark & \checkmark & \checkmark & 0.23 & 0.48 & 82.7 & 89.3 & 92.3 & 22.9 \\ 
\addlinespace
 & \textbf{I-HR iter 0} & \textbf{Iter 0} & $\times$ & $\times$ & \checkmark & $\times$ & 0.55 & 0.78 & 62.4 & 79.9 & 87.1 & 2.0 \\ 
\midrule
\multirow{5}{*}[-5pt]{P-GT} & \multirow{3}{*}{\textbf{Fixed iter}} & \textbf{Iter 0} & \checkmark & \checkmark & $\times$ & $\times$ & 0.53 & 0.72 & 61.1 & 79.7 & 87.5 & 0.5 \\
 &  & \textbf{Iter 1} & $\times$ & $\times$ & \checkmark & $\times$ & 0.50 & 0.72 & 64.6 & 81.7 & 88.6 & 2.5 \\
 &  & \textbf{Iter 2} & $\times$ & $\times$ & $\times$ & \checkmark & 0.27 & 0.48 & 78.3 & 87.8 & 91.5 & 25.4 \\ 
\addlinespace
 & \textbf{All-always} & \textbf{Iter 0} & \checkmark & \checkmark & \checkmark & \checkmark & 0.21 & 0.46 & 83.0 & 88.8 & 91.5 & 22.9 \\ 
\addlinespace
 & \textbf{I-HR iter 0} & \textbf{Iter 0} & $\times$ & $\times$ & \checkmark & $\times$ & 0.53 & 0.75 & 65.1 & 82.1 & 88.5 & 2.0 \\ 
%\midrule
%\multirow{3}{*}[-5pt]{$\times$} & \textbf{US-only} & $\times$ & \checkmark & $\times$ & $\times$ & $\times$ & 0.70 & 0.91 & 49.9 & 66.3 & 73.8 & 0.02 \\ 
%\addlinespace
% & \textbf{ToF-only} & $\times$ & $\times$ & $\times$ & $\times$ & \checkmark & 0.25 & 0.56 & 82.6 & 88.7 & 91.3 & 20.9 \\ 
%\addlinespace
% & \textbf{Dummy} & $\times$ & $\times$ & $\times$ & $\times$ & $\times$ & 0.59 & 0.79 & 39.9 & 80.0 & 86.6 & $\times$ \\
\bottomrule
\end{tabular}
} % \resizebox
\end{table*}

%% file: 5_results.tex
\section{Results} \label{sec:results}

\subsection{Recurrent Transformer Evaluation} \label{sec:results_transformer}

\textbf{Setup:} We evaluate our system on the NYUv2 dataset~\cite{silberman2012indoor}.
Results are summarized in Table~\ref{tab:nyuv2_energy_results}, reporting MAE, and RMSE in meters, together with the $\delta_1$, $\delta_2$, and $\delta_3$ accuracy metrics.
Each $\delta_k$ measures the percentage of pixels for which the predicted depth $\hat{D}$ satisfies
$\max\!\left(\frac{\hat{D}}{D_{GT}}, \frac{D_{GT}}{\hat{D}}\right) < 1.25^k$ 
where $D_{GT}$ denotes the ground truth depth and $k \in \{1,2,3\}$.

We compare our progressive sensor fusion scheme (fixed iter), introduced in Section~\ref{sec:transformer_arch}, against several baselines: \textit{i}) \emph{All-always}, where all sensor modalities are processed jointly in a single iteration; \textit{ii}) \emph{I-HR Iter 0}, which uses only the 160$\times$\SI{160}{\pixel} high-resolution camera at iteration~0; and \textit{iii}) \emph{Dummy} predictor, which always predicts the per-pixel mean of the NYUv2 training set.
Additionally, we report two baselines requiring no network training/inference: \emph{US} and \emph{ToF} depth measurements, rescaled to $160\times160$ resolution to match the ground truth.
These experiments do not use temporal memory tokens (Section~\ref{subsec:temporal_tokens}) or the confidence estimation network (Section~\ref{subsec:confidence_estimation}).
We first assess the impact of the use of pseudo-labels in training on depth prediction error, and then analyze the accuracy-energy trade-off enabled by progressive sensor fusion.

\subsubsection{Ground truth vs. pseudo-labels}

Table~\ref{tab:nyuv2_energy_results} reports the depth reconstruction metrics for models trained using either the GT from the Kinect, which suffers from missing measurements  as explained in Section~\ref{subsec:pseudo_labels}, and those trained with our dense pseudo-labels P-GT.
Despite noisy, incomplete supervision, models trained with Kinect GT still achieve competitive performance across all configurations, demonstrating the robustness of our fusion architecture to degraded training data.
As a reference baseline, we employ a dummy predictor that outputs the per-pixel mean depth of the training set, achieving \SI{0.59}{\meter} MAE, \SI{0.79}{\meter} RMSE, and 39.9\% $\delta_1$, i.e., substantially underperforming all of our models in both training versions, GT and P-GT.
For the progressive sensor fusion scheme, training with P-GT reduces MAE to \SI{0.27}{\meter}–\SI{0.53}{\meter}, compared to \SI{0.32}{\meter}–\SI{0.58}{\meter} when training with GT.
The largest improvement occurs at iteration~1, where predictions rely primarily on visual cues before ToF measurements are incorporated, highlighting the importance of clean ground truth data for training.
When all sensors are jointly processed (\textbf{All-always}), training with P-GT yields an 8.6\% reduction in MAE and improves $\delta_1$ by 2.7\%.
This indicates that higher-quality visual features facilitate more effective multimodal fusion, enabling the transformer to better exploit depth inputs and produce sharper predictions.
Overall, these results validate the use of pseudo-labels for training, as they improve visual feature learning and enhance multimodal depth estimation.
We adopt P-GT for training in all subsequent experiments.

\subsubsection{Energy-Accuracy trade-off}

Focusing on models trained with pseudo-labels, Table~\ref{tab:nyuv2_energy_results} reports depth estimation errors and energy consumption per round (energy/round), accounting for both sensor acquisition and GAP9 processing operating at its maximum frequency of \SI{370}{\mega\hertz} at \SI{0.8}{\volt}.
The results quantify the impact of incremental sensor utilization on both depth accuracy and energy consumption.
Each iteration introduces a new sensor modality at increasing energy cost: iteration 0 (camera and ultrasound) achieves \SI{0.53}{\meter} MAE at only \SI{0.5}{\milli\joule\per\round}; iteration 1 (adding the high-resolution camera) reduces MAE to \SI{0.50}{\meter} at \SI{2.5}{\milli\joule\per\round}; and iteration 2 (adding the ToF sensor) reaches \SI{0.27}{\meter} MAE at \SI{25.4}{\milli\joule\per\round}.
Overall, incremental sensor utilization improves accuracy by 49\% from iteration 0 to 2 at the cost of a $\sim$50$\times$ increase in energy, with the largest single gain occurring at iteration 2, where the introduction of ToF depth reduces MAE by 46\% (from \SI{0.50}{\meter} to \SI{0.27}{\meter}).
This improvement is consistently reflected across RMSE and $\delta$ metrics: RMSE drops by 33\% between iterations 1 and 2, while $\delta_1$, $\delta_2$, and $\delta_3$ increase by 13.7\%, 6.1\%, and 3.9\%, respectively, confirming that each additional sensor modality contributes measurably to depth estimation quality.
The larger gains in RMSE and $\delta_1$ at iteration 2 indicate that ToF measurements primarily reduce large depth errors and suppress outliers.

\input{table/2_2}

Single-sensor baselines in Table~\ref{tab:nyuv2_energy_results_baselines} further isolate each modality's individual contribution, without any neural network processing.
The raw US measurement provides the cheapest depth reading (\SI{0.02}{\milli\joule}) but, being a single scalar, yields the poorest accuracy (\SI{0.70}{\meter} MAE, 49.9\% $\delta_1$). 
The raw ToF $8\times8$ achieves the best single-sensor accuracy (\SI{0.25}{\meter} MAE, 82.6\% $\delta_1$) but at the highest energy cost (\SI{20.9}{\milli\joule}). 
These baselines highlight the complementary nature of our sensor suite: the US sensor provides a low-cost global depth anchor, the camera contributes dense spatial structure that the US and ToF lack, and the ToF delivers accurate metric depth that the camera cannot infer.

Using only the high-resolution image (\textbf{I-HR Iter 0}) yields \SI{0.53}{\meter} MAE and 65.1\% $\delta_1$, demonstrating that a single visual sensor is insufficient for accurate depth estimation on all the frames within a sub-\SI{100}{\milli\watt} power budget.
Progressive sensor fusion closes this accuracy gap, reducing MAE by 49\% and improving $\delta_1$ by 13.2\%, by incrementally incorporating complementary depth measurements only when needed.
When all sensors are processed jointly (\textbf{All-always}), inference achieves the upper-bound performance of \SI{0.21}{\meter} MAE and 83.0\% $\delta_1$, but at an energy cost of \SI{22.9}{\milli\joule\per\round}, dominated by the ToF sensor (\SI{20.9}{\milli\joule}).
Crucially, the progressive scheme reaches within \SI{0.06}{\meter} MAE and 4.7\% $\delta_1$ of this upper bound while enabling early termination for simple scenes, bypassing the energy-expensive ToF acquisition entirely, thus optimizing the energy-accuracy trade-off by paying the full sensor cost only for complex inputs.

\begin{table}[t!]
    \centering
    \caption{Results on NYUv2 using temporal memory tokens ($\mathbf{T}$), and studying the accuracy-energy trade-off with different p$_{\text{term}}$.}
    \label{tab:results_temporal}
    \resizebox{\columnwidth}{!}{%
    \begin{tabular}{>{\bfseries}c >{\bfseries}c ccccccc}
    \toprule
    \multirow{2}{*}{$\mathbf{T}$} &
    \multirow{2}{*}{p$_{\text{term}}$} &
    \multicolumn{2}{c}{\textbf{Error} $\downarrow$ [\si{\meter}]} &
    \multicolumn{3}{c}{\textbf{Accuracy} $\uparrow$ [\%]} &
    \multirow{2}{*}{\shortstack{\textbf{Energy/round}\\$\downarrow$ [\si{\milli\joule\per\round}]}} \\
    \cmidrule(lr){3-4}\cmidrule(lr){5-7}
     & & \textbf{MAE} & \textbf{RMSE} & \textbf{$\delta_1$} & \textbf{$\delta_2$} & \textbf{$\delta_3$} & \\
    \midrule

    0  & \textbf{0}    & 0.27 & 0.40 & 77.8 & 88.0 & 91.8 & 25.4\\
    0  & \textbf{1/32} & 0.27 & 0.41 & 77.3 & 87.7 & 91.6 & 23.9\\
    0  & \textbf{1/16} & 0.28 & 0.42 & 76.5 & 87.4 & 91.5 & 22.5\\
    0  & \textbf{1/8}  & 0.30 & 0.45 & 74.6 & 86.6 & 91.1 & 19.9\\
    0  & \textbf{1/4}  & 0.34 & 0.49 & 71.5 & 85.3 & 90.4 & 14.8\\
    0  & \textbf{1/2}  & 0.41 & 0.56 & 66.5 & 83.2 & 89.3 & 7.3\\
    0  & \textbf{3/4}  & 0.45 & 0.60 & 63.3 & 81.8 & 88.6 & 2.4 \\

    \midrule
    1  & \textbf{0}    & 0.24 & 0.39 & 80.3 & 88.9 & 92.2 & 25.4\\
    1  & \textbf{1/32} & 0.24 & 0.39 & 80.2 & 88.8 & 92.2 & 23.9\\
    1  & \textbf{1/16} & 0.24 & 0.39 & 80.1 & 88.8 & 92.1 & 22.5\\
    1  & \textbf{1/8}  & 0.24 & 0.39 & 79.7 & 88.6 & 92.1 & 19.9\\
    1  & \textbf{1/4}  & 0.25 & 0.40 & 78.8 & 88.2 & 91.9 & 14.9\\
    1  & \textbf{1/2}  & 0.29 & 0.44 & 75.6 & 86.9 & 91.2 & 7.3\\
    1  & \textbf{3/4}  & 0.38 & 0.53 & 68.2 & 83.9 & 89.6 & 2.4 \\

    \midrule
    2  & \textbf{0}    & 0.22 & 0.37 & 82.0 & 89.6 & 92.6 & 25.5\\
    2  & \textbf{1/32} & 0.22 & 0.37 & 81.9 & 89.5 & 92.6 & 24.0\\
    2  & \textbf{1/16} & 0.22 & 0.37 & 81.7 & 89.5 & 92.5 & 22.5\\
    2  & \textbf{1/8}  & 0.23 & 0.38 & 81.4 & 89.3 & 92.4 & 19.9\\
    2  & \textbf{1/4}  & 0.24 & 0.39 & 80.5 & 88.9 & 92.2 & 14.9\\
    2  & \textbf{1/2}  & 0.28 & 0.43 & 77.3 & 87.6 & 91.5 & 7.3\\
    2  & \textbf{3/4}  & 0.38 & 0.53 & 69.3 & 84.4 & 89.8 & 2.5\\

    \midrule
    16 & \textbf{0}    & 0.23 & 0.38 & 80.6 & 89.0 & 92.4 & 25.8\\
    16 & \textbf{1/32} & 0.23 & 0.38 & 80.5 & 89.0 & 92.3 & 24.3\\
    16 & \textbf{1/16} & 0.24 & 0.39 & 80.3 & 88.9 & 92.2 & 22.9\\
    16 & \textbf{1/8}  & 0.24 & 0.39 & 80.0 & 88.7 & 92.2 & 20.3\\
    16 & \textbf{1/4}  & 0.25 & 0.40 & 79.0 & 88.3 & 92.0 & 15.2\\
    16 & \textbf{1/2}  & 0.29 & 0.44 & 75.4 & 86.8 & 91.1 & 7.6\\
    16 & \textbf{3/4}  & 0.38 & 0.53 & 68.0 & 83.7 & 89.7 & 2.7\\
    \bottomrule
    \end{tabular}
    }
\end{table}

\subsection{Recurrent Transformer Evaluation with Temporal Memory Tokens}

\textbf{Setup:} Table~\ref{tab:results_temporal} evaluates the proposed recurrent transformer with temporal memory tokens on the NYUv2 dataset, with all models trained using pseudo-label supervision.
We first analyze the impact of temporal memory on depth estimation accuracy.
We then study the accuracy-energy trade-off with adaptive sensor fusion.
We compare against a \emph{random} baseline, where inference is terminated at each iteration with a fixed probability $p_{\text{term}} \in [0,1]$, independently of the input.
By varying $p_{\text{term}} \in \{0, 1/32, 1/16, 1/8, 1/4, 1/2, 3/4\}$, we characterize the expected depth error and energy consumption, spanning behaviors from no early termination ($p_{\text{term}}=0$) to aggressive early termination ($p_{\text{term}}=3/4$).
We compare this probabilistic policy with adaptive sensor fusion driven by the proposed confidence estimation network.
Finally, we validate the complete system on an in-field dataset collected with our multimodal active ranging deck.

\subsubsection{Temporal memory tokens} \label{subsec:results_temp_tokens}

In Table~\ref{tab:results_temporal} we analyze the impact of temporal token propagation on depth estimation accuracy and temporal consistency, by varying the number of temporal memory tokens $N_t \in {0,1,2,16}$ without early termination ($p_{\text{term}}=0$).
The stateless baseline ($N_t=0$) processes each round independently, discarding all information from previous frames. 
Introducing temporal memory consistently improves performance: with $N_t=2$, the model reduces MAE from \SI{0.27}{\meter} to \SI{0.22}{\meter} (18.5\%) and increases $\delta_1$ from 77.8\% to 82.0\%, with a negligible energy overhead of \SI{0.1}{\milli\joule} per round. 
This improvement demonstrates that propagating even a small amount of context across rounds enables the transformer to leverage temporal correlation between consecutive frames, producing sharper, more consistent depth predictions.

Beyond $N_t=2$, performance saturates: $N_t=16$ slightly degrades accuracy (e.g., $\delta_1$ 80.6\% vs.\ 82.0\%) while increasing energy consumption.
We conducted more tests with $N_t \in \{4,8,32,64\}$, which confirm the plateau in depth estimation error, and are omitted for brevity.
Based on these results, we adopt $N_t=2$ in all subsequent experiments, which introduces a negligible energy/round overhead (+\SI{0.1}{\milli\joule}), and overhead of $\sim$\SI{1}{\kilo\nothing} parameters and results in a total model size of \SI{688}{\kilo\nothing} parameters including the confidence estimation network (\SI{4}{\kilo\nothing} parameters).

\subsubsection{Energy-accuracy trade-off with random early termination} \label{subsec:results_stochastic}

Table~\ref{tab:results_temporal} summarizes the accuracy-energy trade-off obtained with the baseline random early termination.
We focus on the $N_t=2$ configuration.
Without early termination ($p_{\text{term}}=0$), the model always executes all iterations, achieving the highest accuracy at an energy cost of approximately \SI{24}{\milli\joule} per round, dominated by ToF acquisition and GAP9 processing.
Low termination probabilities ($p_{\text{term}}=1/32$ and $1/16$) reduce energy consumption to \SI{24.0}{\nothing} and \SI{22.5}{\milli\joule\per\round}, respectively, with no degradation in depth accuracy.
As $p_{\text{term}}$ increases, energy savings become more pronounced at the cost of gradually increasing error.
For instance, $p_{\text{term}}=1/4$ increases MAE by only \SI{0.02}{\meter} while reducing energy to \SI{14.9}{\milli\joule\per\round}.
At $p_{\text{term}}=1/2$, the model achieves \SI{0.28}{\meter} MAE with \SI{7.3}{\milli\joule} per round, while at $p_{\text{term}}=3/4$ energy drops to \SI{2.5}{\milli\joule\per\round} with a higher error of \SI{0.38}{\meter}.
Overall, random early termination enables a wide and tunable accuracy-energy operating range.

Comparing $N_t=0$ and $N_t=2$, temporal token propagation consistently improves the accuracy-energy trade-off across all $p_{\text{term}}$ values.
The benefit is most pronounced under aggressive early termination, where fewer sensor modalities are processed per round: at $p_{\text{term}}=1/2$, the temporal model ($N_t=2$) achieves \SI{0.28}{\meter} MAE compared to \SI{0.41}{\meter} for the stateless baseline, a 32\% error reduction at identical energy consumption (\SI{7.3}{\milli\joule} per round).
These results highlight that temporal memory enables earlier termination while preserving accuracy:  they allow the transformer to reuse prior geometric information when an iteration terminates early and compensating for missing sensor acquisitions, ultimately yielding superior accuracy-energy trade-offs.

\begin{figure*}[t]
\centering
\includegraphics[width=1\textwidth]{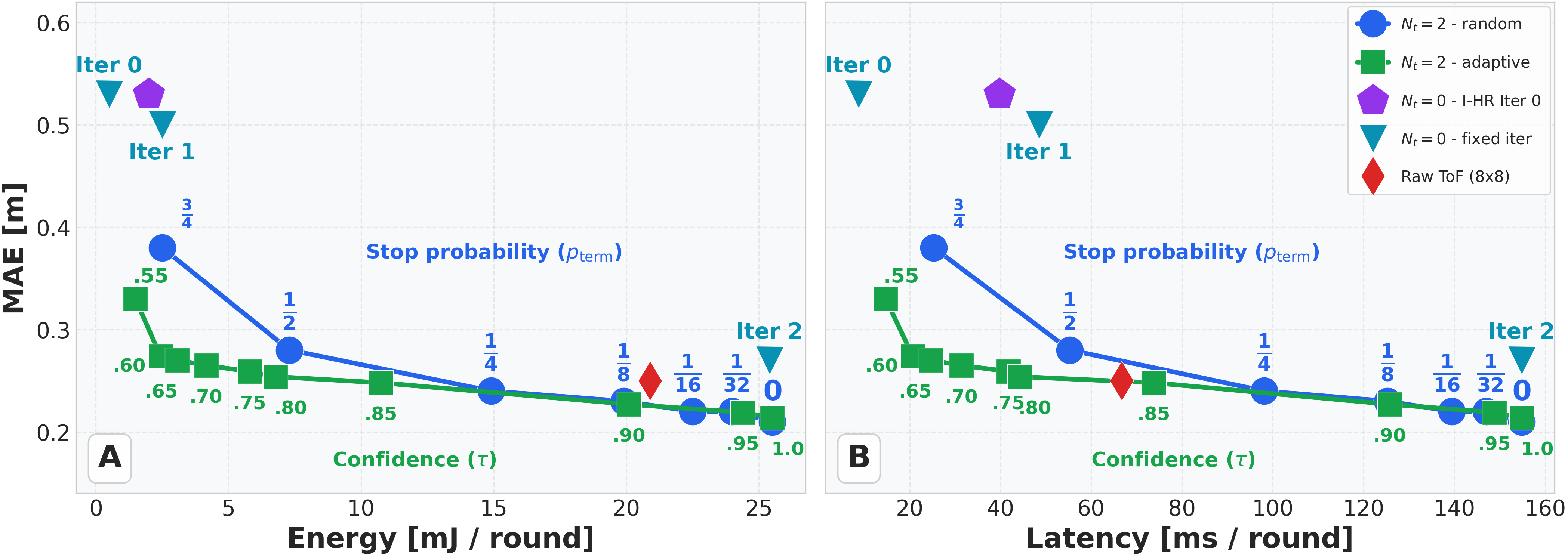}
\caption{Average energy consumption (A) and latency (B) vs. MAE on NYUv2 test set, considering computation and sensor acquisition.
Numbers on the blue points indicate the $p_{\text{term}}$, while numbers on the green squares indicate the confidence threshold $\tau$.
}
\label{fig:results_plot}
\end{figure*}

\subsubsection{Adaptive sensor fusion with our confidence estimation network}

using the recurrent transformer with temporal memory tokens ($N_t=2$), we evaluate the proposed \emph{adaptive sensor fusion} strategy driven by the confidence estimation network (Section~\ref{subsec:confidence_estimation}) and compare it against the random early termination baseline (Section~\ref{subsec:results_stochastic}).
We sweep the confidence threshold $\tau \in [0.5,1]$.
Fig.~\ref{fig:results_plot} reports the MAE on NYUv2 as a function of the energy consumed per inference round.

At high confidence thresholds ($\tau \geq 0.90$), most scenes are classified as insufficiently confident in early iterations, causing the system to frequently execute all three iterations, achieving accuracy-energy trade-offs comparable to random early termination
For lower thresholds ($\tau \leq 0.85$), the confidence network effectively discriminates between simple and complex scenes: geometrically simple inputs terminate after iteration 0 or 1 using only low-cost sensors (\SI{0.5}{\milli\joule}--\SI{2.5}{\milli\joule}), while complex scenes with unresolved depth structure proceed to iteration 2 to incorporate the ToF sensor (\SI{25.4}{\milli\joule}).
We next examine two representative operating points that highlight the advantage of adaptive sensor fusion over input-agnostic random early termination.
The first maximizes the MAE improvement under a fixed energy budget.
The second minimizes energy consumption at iso-MAE.
At $\tau = 0.60$, the average energy consumption of our adaptive sensor fusion drops to only \SI{2.44}{\milli\joule\per\round}, indicating that the majority of NYUv2 test frames are resolved with one or two iterations.
At this operating point, the system achieves \SI{0.27}{\meter} MAE, corresponding to a 29\% improvement over random termination at the same energy budget.
At $\tau = 0.75$, energy consumption decreases by 60\% compared to random termination at iso-MAE (from \SI{14.9}{\milli\joule\per\round} to \SI{5.8}{\milli\joule\per\round}).
This shows that our confidence network correctly identifies challenging samples, triggering additional iterations to refine the output by incorporating additional sensors.

Overall, these results show that \textbf{adaptive sensor fusion is particularly effective in energy-constrained regimes.
It selectively triggers multimodal sensor fusion only for complex scenes, and it consistently outperforms random early termination, achieving the same MAE with up to 60\% lower energy consumption.}
Simple scenes terminate early using only low-cost sensors (camera and ultrasound, \SI{0.5}{\milli\joule}), while complex scenes trigger the full sensor suite, including the power-hungry ToF (\SI{20.9}{\milli\joule}).

Finally, Fig.~\ref{fig:results_plot} presents a comprehensive comparison of our complete system, integrating confidence-based adaptive sensor fusion with temporal memory tokens.
We compare it against fixed-iteration progressive fusion (fixed iter, Section~\ref{sec:results_transformer}) and single-sensor baselines (raw ToF and US), neither of which employs adaptive or temporal mechanisms.
The comparison reveals two complementary advantages of our approach. 
First, at iso-energy ($\sim$\SI{25.4}{\milli\joule\per\round}), temporal memory tokens improve accuracy by 22\%, reducing MAE from \SI{0.27}{\meter} (fixed progressive fusion, iteration 2, \SI{154.8}{\milli\second\per\round}) to \SI{0.21}{\meter}, as cross-round context propagation enables the transformer to produce sharper predictions without additional sensor cost. 
Second, at iso-accuracy ($\text{MAE}=\SI{0.27}{\meter}$), the confidence-based gating allows our system to reach the same error as the full fixed pipeline while consuming only \SI{2.44}{\milli\joule\per\round} ($\tau=0.60$), a 90\% energy reduction and $7.5\times$ latency improvement (from \SI{154.8}{\milli\second} to \SI{20.7}{\milli\second\per\round}), retaining most of the full-pipeline $\delta_1$ accuracy (77.1\% vs. 82.0\%).
Fig.~\ref{fig:results_plot}-B further illustrates this latency advantage: adaptive gating spans a continuous throughput range from \SI{6}{frame/\second} (full pipeline, $\tau=1.0$) to \SI{48.3}{frame/\second} ($\tau=0.60$), compared to random termination which achieves similar throughput only at significantly higher MAE, confirming that scene-aware gating reduces not only energy but also per-round latency by avoiding unnecessary iterations.

Single-sensor baselines bound the operating range: US inference minimizes energy (\SI{0.02}{\milli\joule\per\round}) but yields poor accuracy ($\text{MAE}=\SI{0.70}{\meter}$), while ToF inference achieves $\text{MAE}=\SI{0.25}{\meter}$ at \SI{20.9}{\milli\joule\per\round} with a maximum output data rate of \SI{15}{\hertz}. 
At comparable energy ($\tau=0.85$, \SI{10.8}{\milli\joule\per\round}), our system matches ToF accuracy while consuming 49\% less energy, demonstrating that adaptive multimodal fusion outperforms even dedicated depth sensors in terms of energy efficiency.
Taken together, these results confirm that \textbf{combining temporal memory tokens with confidence-based adaptive sensor fusion fundamentally shifts the accuracy-energy Pareto frontier: our system retains over 94\% of full-pipeline $\delta_1$ while reducing energy by up to 90\% and latency by $7.5\times$, enabling scene-adaptive multimodal depth estimation within ultra-low-power constraints.}

\subsubsection{In-field evaluation}

to validate our system on a real-world use case, we evaluate our recurrent transformer, using temporal memory tokens and the adaptive sensor fusion, on an in-field dataset of approximately 1000 frames collected with our custom multimodal active ranging deck, integrating a low-resolution camera, an US depth sensor, and an $8\times8$ ToF depth array.
This experiment is subject to a domain shift gap: the model is trained solely on NYUv2, where US and ToF measurements are synthetically derived from Kinect RGB-D data, and then deployed zero-shot on real hardware exhibiting different noise, sparsity, and failure modes.
Table~\ref{tab:infield_results} reports the resulting zero-shot performance across different confidence thresholds $\tau$, where Dummy is a baseline that always predicts the per-pixel mean of the training set.

Crucially, the in-field results exhibit the same monotonic improvement pattern observed on the NYUv2 test set: as the confidence threshold increases from $\tau = 0.4$ to $\tau = 1.0$, MAE decreases from \SI{0.80}{\meter} to \SI{0.63}{\meter} (21\% reduction), RMSE from \SI{1.23}{\meter} to \SI{0.99}{\meter} (20\% reduction), while $\delta_2$ improves from 73.3\% to 84.4\% (+11.1\%) and $\delta_3$ from 86.1\% to 91.3\% (+5.2\%).
While absolute errors are higher than on NYUv2 due to differences between the Kinect training data and our actual sensor characteristics, including different camera matrices, noise profiles, and the ToF sensor's narrower field of view, the multi-sensor fusion architecture successfully generalizes beyond the training distribution.
The consistent accuracy-energy trade-off across thresholds further validates that the FFT-based confidence network correctly identifies when additional refinement yields accuracy gains, even under this domain shift.
Operators can select confidence thresholds based on the operative requirements, prioritizing energy efficiency for extended operating time ($\tau \leq 0.6$) or accuracy for safety-critical operations ($\tau \geq 0.9$), with assurance that the expected trade-offs will hold in deployment.

\begin{table}[t]
\centering
\large
\caption{In-field evaluation of our recurrent transformer for multimodal sensor fusion, exploiting temporal memory tokens ($N_t = 2$) and adaptive sensor fusion.}
\label{tab:infield_results}
\resizebox{\columnwidth}{!}{%
\begin{tabular}{c c c c c c c c}
\toprule
    \multirow{2}{*}{\textbf{Model}} &
    \multirow{2}{*}{$\tau$} &
    \multirow{2}{*}{\shortstack{\textbf{MAE $\boldsymbol{\downarrow}$}\\{[\si{\meter}]}}} &
    \multirow{2}{*}{\shortstack{\textbf{RMSE $\boldsymbol{\downarrow}$}\\{[\si{\meter}]}}} &
    \multirow{2}{*}{\shortstack{\textbf{$\delta_1$ $\boldsymbol{\uparrow}$}\\{[\%]}}} &
    \multirow{2}{*}{\shortstack{\textbf{$\delta_2$ $\boldsymbol{\uparrow}$}\\{[\%]}}} &
    \multirow{2}{*}{\shortstack{\textbf{$\delta_3$ $\boldsymbol{\uparrow}$}\\{[\%]}}} &
    \multirow{2}{*}{\shortstack{\textbf{Energy/round $\boldsymbol{\downarrow}$}\\{[\si{\milli\joule\per\round}]}}} \\
    &&&&&&&\\
\midrule
Our & 0.4 & 0.80 & 1.23 & 58.4 & 73.3 & 86.1 & 0.5 \\
Our & 0.6 & 0.74 & 1.16 & 63.8 & 76.9 & 87.4 & 2.2 \\
Our & 0.8 & 0.73 & 1.08 & 58.2 & 78.3 & 88.9 & 12.6 \\
Our & 0.9 & 0.67 & 1.02 & 61.2 & 82.0 & 90.8 & 19.7 \\
Our & 1.0 & 0.63 & 0.99 & 64.0 & 84.4 & 91.3 & 25.5 \\
\midrule
Dummy & --- & 1.18 & 1.50 & 9.9 & 57.0 & 81.0 & --- \\
%\midrule
%$\times$ & 0.63 & 0.99 & 64.0 & 84.4 & 91.3 & 20.9 \\
%$\times$ & 0.63 & 0.99 & 64.0 & 84.4 & 91.3 & 0.02 \\
%$\times$ & 0.63 & 0.99 & 64.0 & 84.4 & 91.3 & $\times$ \\
\bottomrule
\end{tabular}
}
\end{table}

\subsubsection{SoA comparison and Discussion}

overall, our multimodal transformer achieves up to \SI{0.21}{\meter} MAE and \SI{0.37}{\meter} RMSE with \SI{688}{\kilo\nothing} parameters at $<\SI{0.4}{\watt}$.
When the adaptive confidence-based gating is disabled, and all three iterations are always executed (i.e., the non-adaptive configuration on the same hardware), the system achieves 82.0\% $\delta_1$ at \SI{25.5}{\milli\joule} per round.
Enabling the adaptive gating ($\tau=0.60$) retains \SI{0.27}{\meter} MAE while consuming only \SI{2.44}{\milli\joule} per round, a $10\times$ energy reduction, demonstrating that the adaptive mechanism is the key enabler for operating within the sub-\SI{100}{\milli\watt} power envelope on battery-powered IoT and cyber-physical systems.

Compared to Wang et al.~\cite{wang2020mobiledepth}, which achieves \SI{0.40}{\meter} MAE with $\sim$\SI{6}{\mega\nothing} parameters, we reduce MAE by $\sim$50\% while using 9$\times$ fewer parameters.
These gains stem from our progressive multimodal fusion, which supplements visual features with direct depth measurements from the ToF array and ultrasound ranger, rather than relying solely on learned monocular features from a larger network.

Nadalini et al.~\cite{nadalini2025multi} is the closest baseline deployed on the same GAP9 hardware platform, targeting monocular depth estimation without adaptive gating or multimodal sensor fusion.
We compare against two configurations:
\textit{(i)} their published results, where the model is pre-trained on TartanAir~\cite{wang2020tartanair} and fine-tuned on-device with ToF supervision, and
\textit{(ii)} the same architecture trained from scratch on NYUv2 without fine-tuning, which we reproduce for fair comparison.

Against configuration~\textit{(i)}, our model improves RMSE by 26\% (\SI{0.37}{\meter} vs. \SI{0.50}{\meter}) and $\delta_1$ by 31.8\% (81.9\% vs.\ 50.1\%), without requiring the costly and time-consuming on-device fine-tuning (\SI{860}{\second}, \SI{204.9}{\joule}) that the configuration demands.
Against configuration~\textit{(ii)}, we improve $\delta_1$ by 32.2\% (81.9\% vs.\ 49.7\%) and RMSE by 22.9\% (\SI{0.37}{\meter} vs. \SI{0.48}{\meter}).
The accuracy gap widens further in our favor, demonstrating that our multimodal fusion and pseudo-label training~\cite{yang2024depth} provide a more effective path to deployment-ready accuracy than pure monocular estimation.
Our 688K-parameter model is 7$\times$ larger than Nadalini et al.'s network, originally designed by Peluso et al. for the more memory-constrained STM32~\cite{peluso2021pypydnet}, but fits within GAP9's memory budget.

From an energy perspective, the two Nadalini configurations present different trade-offs.
Configuration~\textit{(i)} requires on-device fine-tuning with continuous ToF supervision (\SI{860}{\second}, \SI{204.9}{\joule}), a one-time but substantial cost that must be repeated if the deployment environment changes.
Moreover, during inference, their monocular network cannot leverage the ToF sensor, and still requires up to \SI{100}{\milli\watt}
In contrast, our approach activates the ToF array opportunistically: only rounds where the confidence network requests additional refinement (iteration 2) incur ToF acquisition energy, while the first two iterations bypass the depth sensor entirely.
At $\tau = 0.6$, we achieve $\delta_1=77.1\%$ within only \SI{2.4}{\milli\joule} per round, amortizing sensor costs across the majority of rounds that use only early iterations, resulting in an average power consumption of up to $\sim$\SI{110}{\milli\watt} through the entire NYUv2 test set at a \SI{48.3}{round/\second} inference frequency.
Configuration~\textit{(ii)} avoids the fine-tuning overhead entirely but, lacking any depth sensor input, achieves substantially lower accuracy (49.7\% $\delta_1$ vs.\ our 81.9\%), still requiring up to \SI{100}{\milli\watt} for the processing alone.
Our architecture thus occupies a favorable middle ground: no fine-tuning cost, optional ToF activation governed by scene complexity, leveraged for subsequent rounds, and higher accuracy than both Nadalini et al. configurations.

\textbf{Discussion.}
From a computational and memory efficiency perspective, the \texttt{int8}-quantized model occupies only \SI{688}{\kilo\byte} of GAP9's \SI{1.5}{\mega\byte} L2 SRAM, enabling all model weights to reside on-chip and avoiding costly L3 HyperRAM accesses during inference. 
Convolution-heavy layers are offloaded to the NE16 hardware accelerator, while attention and element-wise operations can be parallelized across the 9 RISC-V cluster cores with SIMD execution.
On the GAP9 running at \SI{370}{\mega\hertz} at \SI{0.8}{\volt}, the full three-iteration pipeline completes in \SI{154.8}{\milli\second} (\SI{6}{frame/\second}), while a single iteration requires only \SI{8.8}{\milli\second} (\SI{114}{frame/\second}).
With adaptive gating ($\tau=0.60$), the average latency drops to \SI{20.7}{\milli\second\per\round} (\SI{48.3}{round/\second}), demonstrating that the adaptive mechanism effectively translates into proportional computational savings on the target hardware.

To illustrate the long-term impact of dynamic early-exit in continuous operation, consider a battery-powered deployment with a typical \SI{1000}{\milli\ampere\hour} Li-Po cell (\SI{13.30}{\kilo\joule} @ \SI{3.7}{\volt}).
The non-adaptive full pipeline (\SI{25.5}{\milli\joule\per\round}) can sustain approximately \SI{522}{\kilo\nothing} depth estimation rounds before battery depletion, whereas the adaptive configuration ($\tau=0.60$, \SI{2.44}{\milli\joule\per\round}) delivers over \SI{5.4}{\mega\nothing} rounds on the same battery, a $10.4\times$ increase.
At a fixed inference rate of \SI{6}{round/\second}, this translates into approximately \SI{24}{\hour} of non-adaptive operation versus over \SI{250}{\hour} (more than 10 days) with adaptive early-exit, demonstrating that the dynamic gating mechanism directly extends operational lifetime by an order of magnitude in battery-powered deployments.

Our proposed adaptive transformer architecture is inherently extensible to additional sensors, and applicable to other perception tasks and embedded platforms.
New sensor streams can be incorporated by adding dedicated encoders (Section~\ref{subsec:encoder}) that project measurements into the shared token space without modifying the recurrent transformer core.
This design can accommodate sensors such as event-based cameras, LiDARs, thermal/infrared cameras, and hyperspectral sensors.
Likewise, the architecture can be adapted to tasks beyond depth estimation by replacing the task-specific decoder (Section~\ref{subsec:decoder}), enabling applications such as semantic segmentation, object detection, and scene classification, just to name a few.
Finally, the compact model footprint (\SI{688}{\kilo\byte} \texttt{int8}) and the use of standard convolution and attention operators make the framework portable across a broad range of ULP embedded platforms beyond GAP9.
To validate this portability, we deployed the transformer architecture on the STM32N6 SoC\cite{11409013}, which features an Arm Cortex-M55 running at \SI{800}{\mega \hertz}.
The deployment achieves an energy consumption of \SI{689.5}{\milli\joule} per round when executing the full \emph{All-always} configuration, demonstrating that the proposed architecture can be deployed on different MCUs without architectural modifications.

%% file: table/2_2.tex
\begin{table}[t!]
    \centering
    \caption{Baselines on the NYUv2 test set. US and ToF report raw sensor measurements rescaled to $160\times160$ resolution; Dummy predicts the per-pixel mean depth of the training set.}
    \label{tab:nyuv2_energy_results_baselines}
    \resizebox{\columnwidth}{!}{%
\begin{tabular}{ccccccc}

\toprule
\multirow{2}{*}{\shortstack{\textbf{Sensor} \\ \textbf{use}}} &
\multirow{2}{*}{\shortstack{\textbf{MAE $\downarrow$} \\ \textbf{[\si{\meter}]}}} &
\multirow{2}{*}{\shortstack{\textbf{RMSE $\downarrow$} \\ \textbf{[\si{\meter}]}}} &
\multirow{2}{*}{\shortstack{\textbf{$\delta_1$ $\uparrow$} \\ \textbf{[\%]}}} &
\multirow{2}{*}{\shortstack{\textbf{$\delta_2$ $\uparrow$} \\ \textbf{[\%]}}} &
\multirow{2}{*}{\shortstack{\textbf{$\delta_3$ $\uparrow$} \\ \textbf{[\%]}}} &
\multirow{2}{*}{\shortstack{\textbf{Energy/round} $\downarrow$ \\ \textbf{[\si{\milli\joule\per\round}]}}} \\
&&&&&&\\
\midrule
\textbf{US} & 0.70 & 0.91 & 49.9 & 66.3 & 73.8 & 0.02 \\
\addlinespace
\textbf{ToF} & 0.25 & 0.56 & 82.6 & 88.7 & 91.3 & 20.9 \\
\addlinespace
\textbf{Dummy} & 0.59 & 0.79 & 39.9 & 80.0 & 86.6 & $\times$ \\
\bottomrule
\end{tabular}%
}
\end{table}

%% file: 6_conclusions.tex
\section{Conclusions} \label{sec:conclusions}

We presented a multimodal depth estimation system for ULP MCUs combining a \SI{688}{\kilo\nothing}-parameter Universal Transformer, progressive sensor fusion, temporal memory tokens, and FFT-based confidence estimation for adaptive sensor fusion.
The system progressively fuses low- and high-resolution images, ultrasound measurements, and 8$\times$8 ToF depth data, while propagating temporal context across inference rounds (+0.1\% tokens).
On NYUv2 dataset, our temporal model reduces the MAE by 22\% compared to the non-temporal version (\SI{0.21}{\meter} vs.\ \SI{0.27}{\meter}), reaching 82.0\% $\delta_1$ accuracy.
Deployed on the GAP9 SoC, the system operates at an average power below \SI{110}{\milli\watt}, with an energy consumption spanning from \SI{0.5}{\milli\joule} to \SI{25.5}{\milli\joule}.
Adaptive sensor fusion reduces inference energy by up to $10\times$ (from \SI{25.5}{\milli\joule} to \SI{2.44}{\milli\joule}) while maintaining \SI{0.27}{\meter} MAE with a threshold $\tau = 0.60$.
Our work provides the basis for future in-field robotic experiments, can serve as a baseline for further sensor integrations, such as ULP event-based cameras, and can serve as a basic building block for more complex multimodal tasks, like visual odometry and simultaneous localization and mapping~\cite{marchei_tinydevo}.

%% file: 7_bios.tex
\begin{IEEEbiography}[{\includegraphics[width=1in,height=1.25in,keepaspectratio,clip]{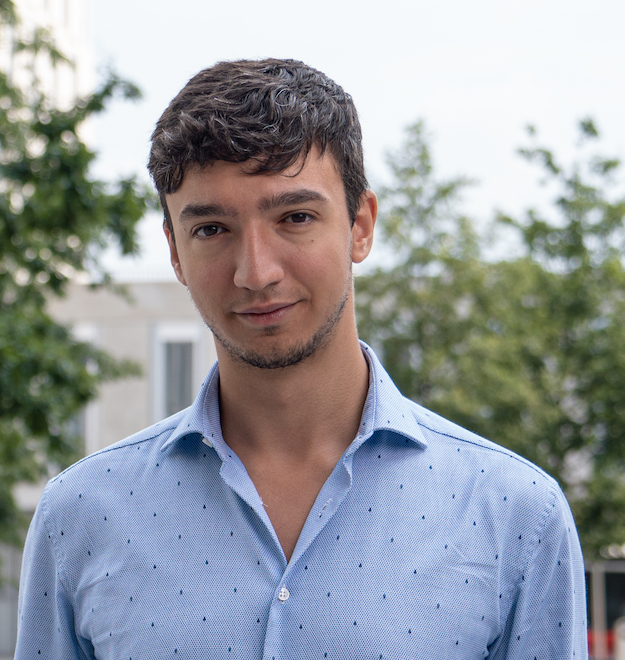}}]
{Luca Crupi~} (Graduate Student Member, IEEE) is a Ph.D. student at the Dalle Molle Institute for Artificial Intelligence (IDSIA, USI-SUPSI) in Lugano, Switzerland. He was part of the XVII of Alta Scuola Politecnica and received an MSc degree in Computer Engineering from Politecnico di Torino in 2022. He was the IT lead at PolitOcean, a student team focusing on the development of a fully autonomous underwater vehicle. His research focuses on developing AI-based autonomous algorithms for pocket-sized robotic platforms integrating multiple sensor streams to perform sensing with neural network-based techniques. His work has resulted in 10+ peer-reviewed publications in international conferences and journals. 
\end{IEEEbiography}

\begin{IEEEbiography}[{\includegraphics[width=1in,height=1.25in,keepaspectratio,clip]{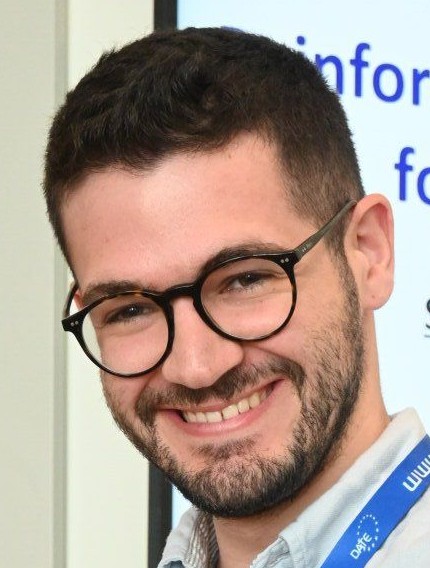}}]{Lorenzo Lamberti~} (Member, IEEE) received his Ph.D. in Electronic Engineering from University of Bologna. 
he is postdoctoral researcher at the Integrated Systems Laboratory (IIS), ETH Zürich, Zürich, Switzerland, and at the Dalle Molle Institute for Artificial Intelligence (IDSIA), SUPSI, Lugano, Switzerland.
He was the technical lead for the winning team of the ``Nanocopter AI Challenge'' at the 2022 IMAV International Conference hosted by TU Delft.
\end{IEEEbiography}

\begin{IEEEbiography}[{\includegraphics[width=1in,height=1.25in,keepaspectratio,clip]{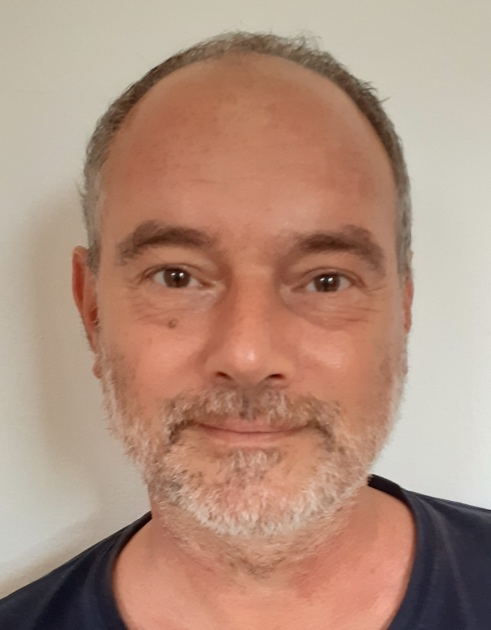}}]
{Giovanni Badaracco~} is a researcher at the Institute of Systems and Applied Electronics (ISEA), Department of Innovative Technologies, SUPSI (University of Applied Sciences and Arts of Southern Switzerland). His work focuses on sensor systems, instrumentation electronics, signal processing, and embedded electronic systems, with applications in industrial sensing and measurement technologies. He has contributed to several applied research projects, including high-precision capacitive sensing and optical speed-sensing systems.
\end{IEEEbiography}

\begin{IEEEbiography}[{\includegraphics[width=1in,height=1.25in,keepaspectratio,clip]{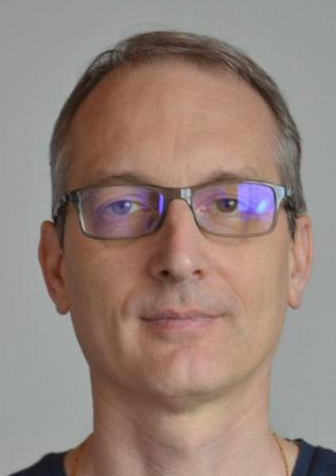}}]
{Daniele Allegri~} (Member, IEEE) is Professor of Programmable and Embedded Microelectronic Systems, Head of the Digital Electronics, Microelectronics and Bioelectronics Scientific Area, and Director of the Institute of Systems and Applied Electronics (ISEA) at the University of Applied Sciences and Arts of Southern Switzerland (SUPSI). He received the Dipl.-Ing. degree in Electronic Engineering from the Swiss Federal Institute of Technology Zurich (ETH Zurich) in 1998 and the Ph.D. degree in Microelectronics from the University of Pavia in 2017. His research spans digital circuits, FPGA-based systems, System-on-Chip architectures, embedded systems, integrated biomedical circuits and systems, analog and mixed-signal integrated circuit design, analog and digital signal processing, machine learning for embedded systems, medical devices, and electrophysiology. He has led and participated in numerous applied research and technology transfer projects in close collaboration with industrial and academic partners. His current research interests include smart sensing, edge computing, high-speed FPGA-based digital signal processing, and the integration of advanced electronic systems for real-world applications.
\end{IEEEbiography}

\vfill\break

\begin{IEEEbiography}[{\includegraphics[width=1in,height=1.25in,keepaspectratio,clip]{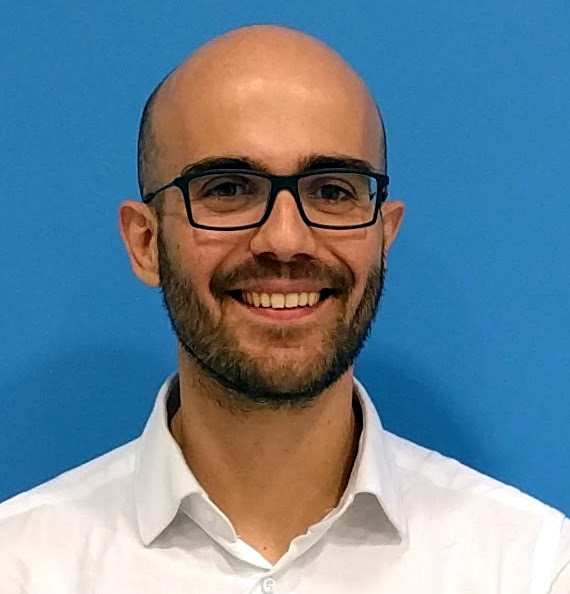}}]{Alessandro Giusti~} is Professor of AI for Autonomous Robotics at the Dalle Molle Institute for Artificial Intelligence (IDSIA, USI-SUPSI) in Lugano, Switzerland. He leads the institute's research area on Autonomous Robotics. His work focuses on Self-Supervised Deep Learning applied to mobile and industrial robotics. He is the author of more than 80 peer-reviewed publications in top conferences and journals and the recipient of several awards, most of which for innovative applications of deep learning to various fields.
\end{IEEEbiography}

\vspace{-12.4cm}

\begin{IEEEbiography}[{\includegraphics[width=1in,height=1.25in,keepaspectratio,clip]{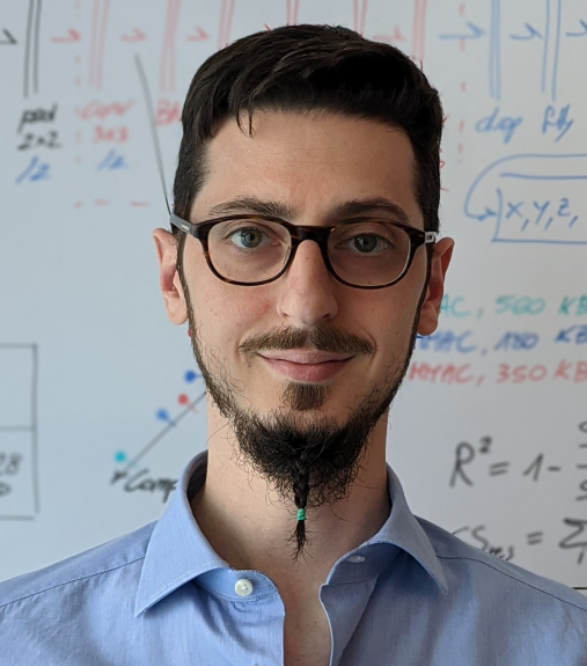}}]{Daniele Palossi~} (Senior Member, IEEE) received his Ph.D. in Information Technology and Electrical Engineering from ETH Z\"urich. He is currently a Senior Researcher and Lecturer at the Dalle Molle Institute for Artificial Intelligence (IDSIA), SUPSI, Lugano, Switzerland, where he leads the NanoRobotics research group, and a Scientific Assistant at the Integrated Systems Laboratory (IIS), ETH Zürich, Zürich, Switzerland. His research stands at the intersection of artificial intelligence, ultra-low-power embedded systems, and miniaturized robotics. His work has resulted in 60+ peer-reviewed publications in international conferences and journals. Dr. Palossi was a recipient of multiple grants from the Swiss National Science Foundation (SNSF); he received the 2nd prize at the Design Contest held at the ACM/IEEE ISLPED'19, several Best Paper Awards, and led the winning team of the first ``Nanocopter AI Challenge'' hosted at the IMAV'22 International Conference.
\end{IEEEbiography}